\documentclass[iop]{emulateapj-rtx4} 
\shortauthors{Sekanina}
\shorttitle{1I/`Oumuamua and Survival of Oort Cloud Comets}
\slugcomment{Version \today }

\newcommand{\gapeq}{$\;$\raisebox{0.3ex}{$>$}\hspace{-0.28cm}\raisebox{-0.75ex}{$\sim$}$\;$}

\newcommand{\Afro}{$\:\!\!A\:\!\!f\!\rho$}
\begin{document}
\title{1I/`Oumuamua and the Problem of Survival of Oort Cloud Comets
  Near the Sun}
\author{Zdenek Sekanina}
\affil{Jet Propulsion Laboratory, California Institute of Technology,
  4800 Oak Grove Drive, Pasadena, CA 91109, U.S.A.}
\email{Zdenek.Sekanina@jpl.nasa.gov{\vspace{0cm}}}

\begin{abstract}
%
%
A 2000--2017 set of long-period comets with high-quality orbits of perihelion
distance $<$1 AU~is~used
to show that the objects that perish shortly before perihelion are nearly
exclusively the Oort Cloud members,
especially those with perihelia within 0.6~AU of the Sun, intrinsically fainter,
and dust poor. Their propensity for disintegration is much higher than
predicted by Bortle's
perihelion survival rule, prompting the author to propose a new synoptic index
to be tested in future prognostication efforts.
By their susceptibility
to demise near the Sun, the nuclei of Oort Cloud comets differ dramatically
from the nuclei of other
long-period comets that almost always survive.  In this scenario,
`Oumuamua --- discovered after
perihelion --- is in all probability a major piece of debris of an interstellar
comet that was bound to
perish near perihelion if it was similar to, though much fainter than,
the~known Oort Cloud comets.  The
nondetection of `Oumuamua by the Spitzer Space Telescope is compatible with
optical data for pancake
shape, but not for cigar shape, with the maximum dimension not exceeding 160~m
(at an 0.1 albedo).
Although the solar radiation pressure induced nongravitational acceleration
requires very high porosity, `Oumuamua's estimated mass is orders of
magnitude greater~than~for~a~cloud of unbound submicron-sized dust grains of
equal cross section.  The acceleration could have displaced `Oumuamua by
250\,000~km in 50~days, scattering
other potential debris over a large volume of space. 
\end{abstract}


\section{Introduction}
My suggestion that 1I/`Oumuamua's parent body had
failed to reach perihelion intact (Sekanina 2019)
was based on the presumed affinity between its physical
behavior and the behavior of intrinsically faint,
dynamically new comets arriving from the Oort Cloud.
The argument further employed the results of a paper
on~the {\small \bf perihelion survival limit} by Bortle (1991),
who noticed that virtually no fainter comets
with perihelion distances less than 0.25~AU have ever
been observed after perihelion, because they perished.
%
Bortle also determined the perihelion-distance dependent
limiting absolute magnitude $H_{\rm surv}$ (normalized
to unit distances from the Sun and Earth) that a comet
has to have in order to survive perihelion passage.
He approximated the observed absolute magnitude with
Vsekhsvyatsky's (1958) quantity $H_{10}$, which assumes
that the visual brightness varies inversely as a fourth
power of heliocentric distance.

Bortle's set of comets with perihelion distance smaller
than 0.5~AU, observed between 1800 and 1988 and fainter
than the survival limit $H_{\rm surv}$, consisted of 23~objects,
divided into three groups.  The most extensive
was the {\it no-survival\/} group with 16~members, while
the {\it survival\/} group (i.e., objects contradicting
the rule) contained a single object.  The intermediate,
{\it unstable-survival\/} group was made up of the
remaining six comets.  This group appears to be rather
heterogeneous; it includes C/1925 X1 (old designation
1926~III, Ensor), which was observed as a headless tail,
the product of preperihelion debris, on the only four
post-perihelion exposures available (Sekanina 1984),
having obviously perished near perihelion.  The group
also includes C/1961~O1 (old designation 1961~V,
Wilson-Hubbard), which was still centrally condensed
on the plates exposed by Tomita (1962) nearly four
months after perihelion --- clearly a survivor.

For some, especially the early 19th-century comets on
Bortle's (1991) list, it is uncertain whether they
were missed after perihelion because of their sudden
fading or because nobody was searching for them, given
that no ephemeris was available.  For example, for comet
C/1801~N1, the first entry on Bortle's list, Kronk
(2003) lists no report on efforts aimed at recovering
the object after perihelion.

One of the 16 nonsurvivors on the list, X/1872 X1,
was an inadvertent product of a search for 3D/Biela
and its existence was so questionable and orbit so
uncertain (Kronk 2003) that it is not even listed in
the Marsden \& Williams (2008) {\it Catalogue of Cometary
Orbits\/}.  There surely is no evidence whatsoever on
its post-perihelion evolution.       

Yet, the most severe problem with the data that Bortle
could do nothing about was the inferior quality of the
orbital elements for nearly all comets in his set.
For fully 20 of the 23 comets, or 87 percent, only
parabolic elements were available.  This explains
why Bortle said his results applied to {\it long-period\/}
comets, a very broad category.  It included members
of the Kreutz sungrazing system on the one hand and
Oort Cloud comets on the other, thus encompassing
a range of orbital periods exceeding four orders of
magnitude, from less than 1000~years to more than
10~million~years.

The inferior quality of the orbital elements was the
corollary of the poor quality of astrometry in early
times and of short orbital arcs observed for nearly
all comets on Bortle's list, which in turn were in part
the product of the early, preperihelion termination of
the observations due to the overly diffuse appearance of
the objects.

Nearly thirty years after Bortle's (1991) pioneering
work it is time to revisit the problem of perihelion
survival, to test his conclusions on a set of high-quality
orbital data, with the aim of avoiding the pitfalls
that he could not circumvent.  Of particular importance
is to find out whether all faint long-period comets
with small perihelion distance have the tendency to
perish or whether this propensity is typical for
a certain group or groups.  Because of the presumed
affinity of `Oumuamua to Oort Cloud comets, they are
the primary focus of this study.  

\section{Long-Period Comets With Small Perihelion
 Distances Discovered in 2000--2017}
In the era of the Sky Survey Projects, of which the
most recent and successful example is PanSTARRS,
comets are often discovered at considerably larger
heliocentric distances than ever before.  With the
parallel substantial progress in the imaging and
image-reduction techniques, the orbital elements
for early discovered comets, including their orbital
periods, are determined in most cases with very high
accuracy.

To begin with, I employed the on-line comet catalog
by S.\ Yoshida\footnote{See {\tt
http://www.aerith.net/comet/catalog/index-code.html}.}
to select, from the list of all comets discovered
between 2000 and 2017, those with a perihelion distance
of less than 1~AU, an orbital period greater than
1000~years, and an observed orbital arc of more than
10~days.  The choices for the orbital period and
orbital arc, which may look extreme, are justified
as they secure a broad range of initial conditions.
The orbital-period limit excludes the Kreutz
sungrazers, including C/2011~W3, from the set.

The resulting set includes 60~comets, presented in
Table~1.  The individual columns list, respectively,
the designation; the perihelion time, $t_\pi$ (in TT);
the perihelion distance, $q$ (in AU); the original
barycentric reciprocal semimajor axis (measuring the
{\vspace{-0.04cm}}orbital period), $(1/a)_{\rm orig}$,
and its mean error (in AU$^{-1}$); the observed
orbital arc, $\Delta t_{\rm obs}$, the temporal
distance of the first, $t_{\rm beg}$, and the last,
$t_{\rm last}$, astrometric observations from the
time of perihelion, \mbox{$\Delta t_{\rm beg} =
t_{\rm beg} \!-\! t_\pi$} and \mbox{$\Delta t_{\rm
last} = t_{\rm last} \!-\! t_\pi$} (all in days);
the observed (preperihelion) absolute magnitude,
$H_0$; the absolute magnitude at Bortle's limit of
perihelion survival, $H_{\rm surv}$; and the reference
to the source of the original semimajor axis.

The elements{\vspace{-0.06cm}} $t_\pi$, $q$, and
$(1/a)_{\rm orig}$ were extracted from the circulars
(NK) issued by S.~Nakano,\footnote{See {\tt
http://www.oaa.gr.jp/$\sim$oaacs/nk.htm}.}
primarily because he lists the residuals from
individual astrometric observations that provide
information on the quality of the orbital solution.  For
several remaining comets the elements  were taken from the
{\it Minor Planet Center\/}'s website\footnote{See {\tt
https://minorplanetcenter.net/db\_search}.} (MPC)
or from specific comprehensive investigations mentioned
below.  The information on the observed orbital arc was taken
from the MPC; $t_{\rm beg}$ and $t_{\rm last}$ are
the times of the actual first and last astrometric
observations, respectively, regardless of whether
they were used in the orbit determination.  Time
$t_{\rm beg}$ may precede the time of discovery,
if prediscovery images were subsequently found and measured,
or it may (slightly) lag behind the discovery
time in a case of visual discovery.  The observed
absolute magnitude was extracted from the on-line
catalog of photometric parameters by
A.~Kammerer,\footnote{See {\tt
http://fg-kometen.vdsastro.de/oldause.htm}.} while
the survival limit was derived from Bortle's (1991)
formula.

Inspection of Table 1 shows the presence of 20~Oort Cloud comets
(33~percent of the total), whose orbital periods are nominally
longer than 3~million years; a small group of 5 comets (8~percent)
with the orbital periods of \mbox{150\,000--1\,000\,000}~years; 23~comets
(38~percent) with the orbital periods of \mbox{2000--50\,000}~years;
and 12~comets (20~percent) with the parabolic or poorly determined
elliptic orbits, similar in quality to most entries in Bortle's
(1991) set; these will be all but ignored in~the following.

\begin{table*}
\vspace{-4.2cm}
\hspace{-0.47cm}
\centerline{
\scalebox{1}{
\includegraphics{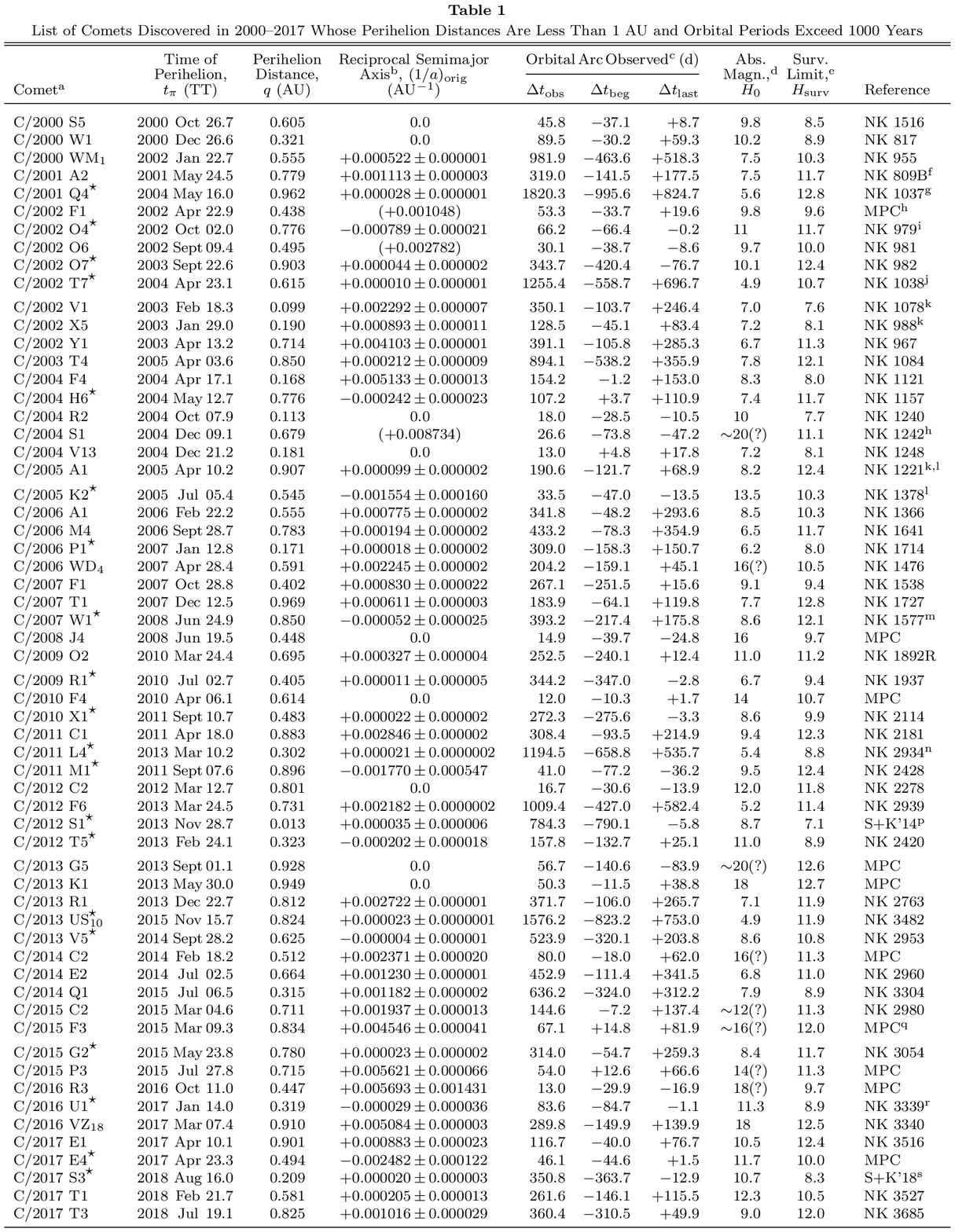}}}
\end{table*}

\section{Signs of Disintegration of a Comet\\Near Perihelion}
The issue now is which of the 60 comets survived perihelion and
which perished.  The decision is a difficult task that requires
examination of the physical behavior of each object to test for
the signs of disintegration.  The symptoms and manifestations are as follows:

(1) Termination of astrometric observations.  If a comet stops suddenly
to be observed for position, this may be an indication that it lost
the nuclear condensation and could no longer be measured.  However,
ground-based astrometry is also terminated when the comet gets too
close to the Sun in the sky; if this is the case, the astrometric
observations should resume after perihelion as soon as the elongation
recovers.  It is thus essential that for comets suspected of perishing near
the Sun the elongation variations be examined during the critical
period of time.  To some extent, this problem is mitigated --- but by no
means eliminated --- by the imaging capabilities of the solar space
observatories, SOHO and STEREO.

(2) Display of progressively increasing, systematic~re\-siduals left
by astrometric observations --- in a period of time shortly before
perihelion --- from an orbital solution that fully satisfied the
observations made only days earlier.  This development is likely to
mark the incipient phase of disintegration of the nucleus into a cloud
of fragments and is typically accompanied by changes in the comet's
appearance.   

\begin{description}
{\scriptsize
\item[\hspace{-0.3cm}]
{\bf Notes to Table 1.} \\[-0.35cm]
\item[\hspace{-0.3cm}]
$^{\rm a}$\,Oort Cloud comets are marked with a star.
\\[-0.47cm]
\item[\hspace{-0.3cm}] 
$^{\rm b}$\,Referred to the barycenter of the Solar System.
\\[-0.47cm]
\item[\hspace{-0.3cm}]
$^{\rm c}$\,Columns give, respectively,
   total period covered by astrometric obser-{\linebreak}
   {\hspace*{-0.7cm}}vations and times of first and last
   observations reckoned from time of{\linebreak}
   {\hspace*{-0.7cm}}perihelion.
\\[-0.47cm]
\item[\hspace{-0.3cm}]
$^{\rm d}$\,Total absolute visual magnitude before
   perihelion (observed or ex-{\linebreak}
   {\hspace*{-0.7cm}}trapolated).
\\[-0.47cm]
\item[\hspace{-0.3cm}]
$^{\rm e}$\,Absolute magnitude at survival limit, given
   by Bortle's formula.
\\[-0.47cm]
\item[\hspace{-0.3cm}]
$^{\rm f}$\,Nongravitational solution.  Comet experienced
   several outbursts and{\linebreak}
   {\hspace*{-0.7cm}}split repeatedly; B was primary nucleus.
   Comet's evolution was inves-{\linebreak}
   {\hspace*{-0.7cm}}tigated in detail by Sekanina et
   al.\,(2002) and by Jehin et al.\,(2002).
\\[-0.47cm]
\item[\hspace{-0.3cm}]
$^{\rm g}$\,Gravitational solution; nongravitational solutions are
   on NK 1126,{\linebreak}
   {\hspace*{-0.7cm}}NK 1265, and NK 1439.
\\[-0.47cm]
\item[\hspace{-0.3cm}]
$^{\rm h}$\,Parenthesized $1/a$ is osculating value;
   no original orbit available.
\\[-0.47cm]
\item[\hspace{-0.3cm}]
$^{\rm i}$\,Orbital variations and related physical
   changes were investigated by{\linebreak}
   {\hspace*{-0.7cm}}Sekanina (2002); see Section 4 for details.
\\[-0.47cm]
\item[\hspace{-0.3cm}]
$^{\rm j}$\,Gravitational solution; nongravitational solution is on
   NK 1438.
\\[-0.47cm]
\item[\hspace{-0.3cm}]
$^{\rm k}$\,Nongravitational solution.
\\[-0.47cm]
\item[\hspace{-0.3cm}]
$^{\rm l}$\,Comet split.
\\[-0.47cm]
\item[\hspace{-0.3cm}]
$\!^{\rm m}\!$\,Gravitational solution; nongravitational solution
   for preperihelion arc{\linebreak}
   {\hspace*{-0.7cm}}of orbit is on NK 1731A.
\\[-0.47cm]
\item[\hspace{-0.3cm}]
$^{\rm n}$\,Orbit determined by T.\ Kobayashi.
\\[-0.47cm]
\item[\hspace{-0.3cm}]
$^{\rm p}$\,Orbit from comprehensive investigation by
   Sekanina \& Kracht (2014).
\\[-0.47cm]
\item[\hspace{-0.3cm}]
$^{\rm q}$\,Member of group with C/1988 A1 (Liller) and
   C/1996 Q1 (Tabur); for{\linebreak}
   {\hspace*{-0.7cm}}details see Sekanina \& Kracht (2016).
\\[-0.47cm]
\item[\hspace{-0.3cm}]
$^{\rm r}$\,Oort Cloud membership somewhat uncertain.
\\[-0.52cm]
\item[\hspace{-0.3cm}]
$^{\rm s}$\,Orbit from comprehensive investigation by
   Sekanina \& Kracht~(2018).}
\\[0cm]
\end{description}

(3) Reports of one or more outbursts in the weeks before perihelion,
followed by a sudden loss of nuclear condensation and subsequent
disappearance of the object.

(4) Presence of a dust tail with a peculiar orientation and one boundary
fairly sharp, indicating a suden dramatic decline in the production of
dust.

(5) Post-perihelion imaging observations at or close to the predicted
location of the comet revealing an object whose appearance differs
dramatically from that of the comet before perihelion, such as
a headless tail, a diffuse and increasingly elongated cloud of debris, etc.

(6) A steep rate of the post-perihelion fading, resulting in major
asymmetry relative to perihelion and accompanied by a drop in the
degree of condensation (DC) in reports by visual observers.

(7) Reports of unsuccessful attempts to detect the comet visually
or with a CCD sensor in an appropriate brightness range, contrary to
expectation based on the preperihelion observations.

The rate at which the disintegration progresses varies from
comet to comet and, if it is relatively slow, a limited number
of post-perihelion observations is not ruled out.  The ultimate
issue of the nature and morphology of the nucleus' debris is a
largely unexplored territory to be addressed in Section~8.

Conservatively, each of the seven warning signs needs to be taken into
consideration before deciding whether a comet survived or perished.
Careful approach is particularly recommended for the Oort Cloud comets,
which are notorious for post-perihelion fading at a rate~that~is always
much steeper than preperihelion brightening~(e.g., Whipple 1978); the
greatly disappointing post-perihelion performance of the would-be
``comet of the century'' C/1973~E1 (Kohoutek), an Oort Cloud object,
will never fade from the witnesses' memory, even though technically
the comet did not perish.

Generally, comets --- from the Oort Cloud or otherwise --- do not always
behave consistently under the~circumstances.  If an object cannot be
positively classified because of inconclusive evidence, only a conditional
judgment is to be made or, in equivocal cases, none at all.{\vspace{0.1cm}}  

\section{Oort Cloud Comets}
For most of the 20 Oort Cloud comets in Table 1 the determination of
the original orbit was straightforward.  In particular, for both
C/2012~S1 (ISON) and C/2017~S3 (PanSTARRS) the tabulated values were
taken from orbital solutions by Sekanina \& Kracht (2014, 2018),
optimized for the purpose of obtaining a reliable original orbit.
The result for C/2001~Q4 was taken from Nakano's gravitational solution
based on a shorter arc, rather than from the subsequent nongravitational
solutions based on longer arcs (NK~1126, NK~1265, and NK~1439), given
the potential problems with extracting an accurate original value once
the nongravitational terms are incorporated into the equations of motion
(Marsden et al.\ 1973).  Similarly, Nakano's gravitational orbits for C/2002~T7
and C/2007~W1 were preferred to his nongravitational orbits on, respectively,
NK~1438 and NK~1731A.  It is a matter of coincidence that for all three
comets the introduction of the nongravitational acceleration turned out
to have at most only a minor effect on the original orbit; there is no
doubt about their having arrived from the Oort Cloud.

While it is well known that the nongravitational forces affect strongly the
orbits of comets with small perihelion distances (Marsden et al.\ 1973, 1978), it
was surprising to find that fully nine of the 20 Oort Cloud comets had the
nominal original orbit hyperbolic.  For three of them --- C/2007~W1 (Boattini),
C/2013~V5~(Oukaimeden),~and C/2016~U1 (NEOWISE)--- the hyperbolic excess
was small, but for C/2004~H6 (SWAN), C/2005~K2 (LINEAR), C/2012~T5 (Bressi),
and C/2017~E4 (Lovejoy)~--- for the last one the distribution of
residuals being~unavailable --- it was in a range of 10--20$\sigma$.  For C/2005~K2,
C/2011~M1 (LINEAR), and C/2017~E4 the hyperbolic excess is understood as part of
the orbital uncertaintes, as these comets were observed for fewer than 50~days.
That {\vspace{-0.025cm}}still leaves two objects, C/2004~H6 and C/2012~T5,
{\vspace{-0.025cm}}\mbox{with\,$(1/a)_{\rm orig}$\,equaling\,$-$0.000242\,$\pm$\,0.000023\,(AU)$^{-1}$\,and}
$-$0.000202\,$\pm$\,0.000018~AU$^{-1}$, respectively, as the least likely
cases of observational errors, given the satisfactory distributions of residuals.
Based on 266 and 603~observations covering the orbital arcs of 107 and
158~days and leaving the mean residuals of $\pm$0$^{\prime\prime\!}$.69 and
$\pm$0$^{\prime\prime\!}$.82, respectively, the two comets have a hyperbolic
excess equivalent to a systematic velocity of nearly 0.5~km~s$^{-1}$ relative
to the Sun at infinity.

Comet C/2004 H6 is exceptional in that it is the only entry in the
data set, for which all astrometry was obtained after perihelion (Table~1).
The object was actually discovered in the UV images, taken with the SWAN
instrument on board the SOHO spacecraft, on 2004 April 29, 13~days
before perihelion, according to three independent reports (Green 2004).
The comet's preperihelion absolute magnitude is extrapolated from the
post-perihelion light curve on the assumption of its perihelion symmetry,
given that there is no evidence of a major preperihelion outburst.

Comet C/2002 O4 (H\"{o}nig) is a difficult case.  Nakano's orbit in Table~1
shows a hyperbolic excess of nealy 40$\sigma$, more than for any other entry.
From the observed evolution of tail orientation, Sekanina (2002) concluded that
the comet was discovered in the course, possibly near the beginning, of an
outburst that engulfed the entire nucleus, resulting in a significant
fraction of its initial mass already lost by the time the event terminated.
The peak post-outburst dust production rate, estimated at about
10$^7$\,g~s$^{-1}$, must have been much higher than the pre-outburst level.
Analyzing Marsden's orbital computations, Sekanina interpreted the gradually
increasing hyperbolic excess as a corollary of progressive crumbling of the
disintegrated nucleus.  The comet's pre-event absolute magnitude could not
be derived from the light curve of the protracted outburst, leaving the
discovery brightness estimate as the best possible source for a crude
approximation.

The roots of the observed hyperbolic excess of the nine comets in Table~1 may
differ from case to case.  For the purpose of the present investigation I
assume that all 20~comets with \mbox{$(1/a)_{\rm orig} < 0.000050$}~AU$^{-1}$
are members of the Oort Cloud.

C/2006 U1 is the most controversial entry in the{\vspace{-0.04cm}} set.
With \mbox{$(1/a)_{\rm orig} = +0.000079$ AU$^{-1}$} at 3$\sigma$,
its orbit is on the inner outskirts of the
Oort Cloud, so that there is a chance, however remote,
that this is not an Oort Cloud comet.  In fact, this
comet may have made a number of revolutions about the
Sun in the past and could have been, at its previous
return to perihelion, diverted toward the Oort Cloud by
the planetary perturbations, a scenario that is plausible
in view of the random nature of the orbital-diffusion
process.  This possibility should be kept in mind, even
though I formally list this object among the Oort Cloud
comets.

Turning now to the examination of the survival status
of the 20 Oort Cloud comets, I first point out that
all those observed astrometrically for at least one to
two months after perihelion are considered survivors
(see Table~1).  The remaining ones are disintegration
suspects, and their status is, one by one, reviewed
below.

To start with, C/2012~T5 was observed for position until
25~days after perihelion.  However, closer inspection
shows that this information is misleading, as the
systematic monitoring of the comet's motion terminated
three weeks {\it before\/} perihelion.  Only an isolated
set of three astrometric positions was obtained after
perihelion on a single night at the Observatoire de Dax
(Code 958), which were not included by Nakano in his
orbital solution.  Ferr\'{\i}n (2014) reported that
the comet's light curve had features typical for the
disintegrating comets.  Kammerer (footnote~4) pointed
out that the comet could not be detected visually after
perihelion.  Neither was it detected in CCD images
15~days after perihelion independently by H.~Sato and
M.~Masek,\footnote{Consult {\tt
https://groups.yahoo.com/neo/groups/comets-ml/
conversations/messages/21071}.},
having been fainter than magnitude~18, more than 9~mag
fainter than three weeks before perihelion; the comet
obviously perished.

The disintegration of C/2002~O7 (LINEAR) is apparent from
a report by Mattiazzo (2003), whose CCD imaging showed the
comet as a headless sunward-pointing tail of debris five
days after perihelion.  According to Tozzi et al.\ (2003),
the comet was fainter than magnitude 20.5, by more than
10~mag compared to its expected brightness, 72~days after
perihelion.

C/2005 K2 split some 2.5 months before perihelion
and four weeks before discovery (Sekanina 2005) and it
flared up about 50~days later (Green 2005).  A CCD image
of the comet's position taken by M.\ Mattiazzo (see
footnote~5, message 8590) 27~days after perihelion
revealed no object brighter than magnitude 16, a clear
sign that the comet did not survive.

A consensus among visual and CCD comet observers was that
C/2009~R1 (McNaught) stopped brightening several weeks before
perihelion.  The comet was very unfavorably located for
observation near perihelion.  J.~\v{C}ern\'y (see
footnote~5, message 16773) reported repeatedly
unsuccessful attempts to detect the comet over the period
of more than three months post-perihelion with a 30-cm
robotic telescope at the Pierre Auger Observatory,
Malarg\"{u}e, Argentina; subsequently he included the
object among the extinct comets in his plot of light
curves.\footnote{See {\tt
http://www.kommet.cz/datas/users/ison+extinct\_1.png}.}
Even though in this case the nature did not cooperate, I
believe the evidence for disintegration is rather strong.
Korsun et al.\ (2012) noted that the spectrum of this
comet at an earlier time displayed an extremely low
continuum, but strong molecular features.

The last astrometric observation of C/2011~M1 was made
more than five weeks before perihelion in spite of favorable
imaging conditions  for another four weeks.  The gap
in the observations was apparently a consequence of the
absence of a nuclear condensation, as implied by the
observations made some 10 or so days before perihelion
(see footnote~5, messages \mbox{17857--17859},~\mbox{17862--17863},
\mbox{17865--17866},~and~\mbox{17870--17872}~that~show~a~fuzzy,~poorly
condensed, ``ghost'' comet.  The object
was apparently not observed after perihelion, in spite
of its increasing elongation to 30$^\circ$ by the time
it was about 1.1~AU from the Sun. The evidence is
compelling enough to deem this Oort Cloud member a nonsurvivor.

The intrinsically faint comet C/2016 U1, observed to
within a day of perihelion passage (14$^\circ$ from the
Sun!), showed no signs of fading.  Starting five days
after perihelion, it appeared for days in the SWAN images
(see footnote~5, message 26174).  It apparently was not
seen from the ground, but it did not reach an elongation
of 35$^\circ$ until nearly four months after perihelion.
All signs point to the comet having survived perihelion
essentially unscathed.

By contrast, C/2017 E4 was a quintessential
example of a comet that perishes near perihelion.  With
the signature of its disintegration overwhelming, it
is sufficient to document the case by referring to a
paper by James (2017) that offers both the dynamical
effect and imaging evidence of its demise.

\begin{table}
\vspace{-4.2cm}
\hspace{4.3cm}
\centerline{
\scalebox{1}{
\includegraphics{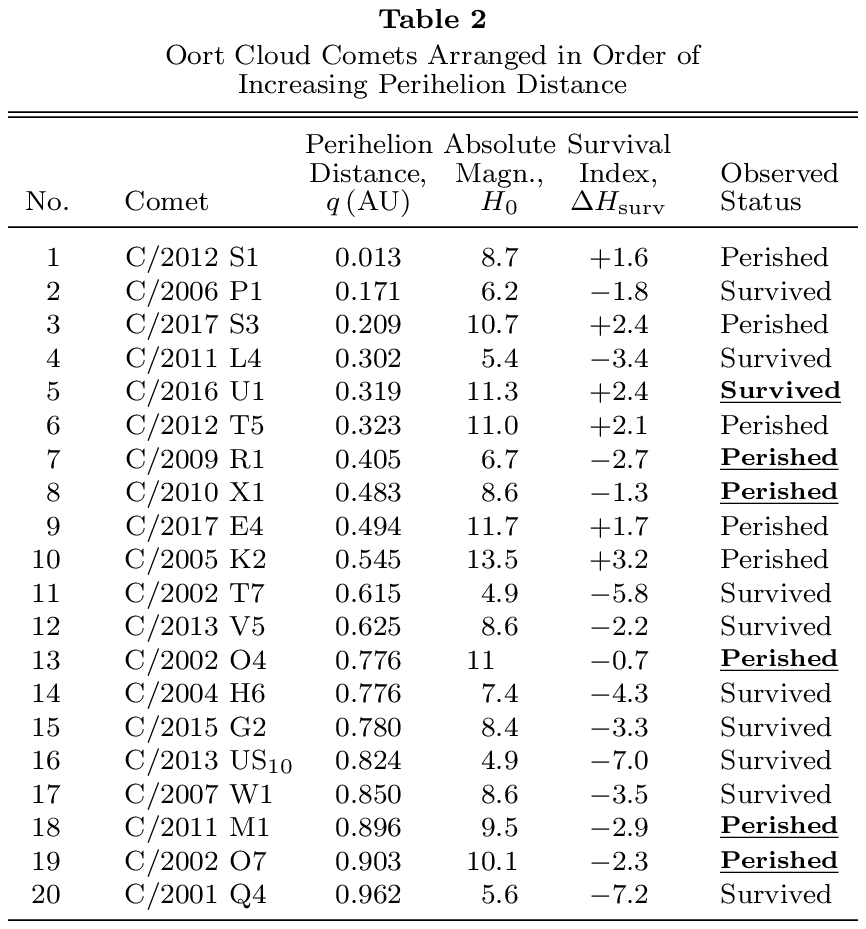}}}
\vspace{-15.8cm}
\end{table}

With strong evidence against survival of the remaining Oort
Cloud comets of interest --- C/2002~O4, C/2010~X1
(Elenin), C/2012~S1 (ISON), and C/2017~S3 --- as presented
elsewhere (e.g., Sekanina 2002; Guido et al.\
2011; Seka\-nina 2011; Ferr\'{\i}n 2014; Knight \& Battams
2014; Seka\-nina \& Kracht 2014; Li \& Jewitt 2015;
Sekanina \& Kracht 2018), it is now possible to address
the two major objectives of this investigation:\ (i)~the
dependence of perihelion survival (or failure to survive)
of an Oort Cloud comet on its perihelion distance and
intrinsic brightness; and (ii)~the degree of conformity
of this relationship to Bortle's (1991) formula for the
survival limit.  In relation to (ii) it should be
remembered that (a)~Bortle did not rule completely out
survival of long-period comets fainter than the limit,
only argued that the ``likelihood of \ldots surviving
perihelion passage becomes drastically reduced'';
(b)~the absolute magnitudes $H_0$ that we use, based on
Kammerer's fit to visual observations, are not identical
to $H_{10}$ that Bortle used following the style of
Vsekhsvyatsky (1958); typically for Oort Cloud comets,
whose preperihelion variation with heliocentric distance $r$
is less steep than $r^{-4}$ (e.g., Whipple 1978),
\mbox{$H_0 < H_{10}$} at \mbox{$r < 1$ AU}
and vice versa.  Since the observed light curve for a
comet with \mbox{$q < 1$ AU} often bridges the point
at \mbox{$r = 1$ AU}, the difference between $H_0$
and $H_{10}$ is usually~insignificant;
and (c)~unlike Bortle, I distinguish between the Oort
Cloud comets and other long-period comets with shorter
orbital periods. 

The Oort Cloud comets from Table~1 are arranged by increasing
perihelion distance in Table~2.  While columns 2--4 are copied
from Table~1, the penultimate column lists a perihelion survival
index, \mbox{$\Delta H_{\rm surv} = H_0 \!-\!  H_{\rm surv}$},
introduced to test the {\it prediction\/} by Bortle's formula.
When \mbox{$\Delta H_{\rm surv} < 0$}, the object is expected
to survive, when \mbox{$\Delta H_{\rm surv} > 0$}, it should
perish.  The last column indicates whether the comet was
{\rm observed\/} to survive or perish, as established in this
section; if defying Bortle's rule, the status is typed in boldface
and is underlined to distinguish these objects from those
that comply with the rule.  The results are astonishing
as they demonstrate that a whopping {\small \bf 50~percent
of~the~Oort~Cloud comets} with perihelia below 1~AU {\small
\bf perish!}  Also, the relationship between the perishing
comets and perihelion distance is stronger than expected:\
70~percent of the objects with \mbox{$q < 0.6$ AU} and
30~percent with \mbox{$0.6 < q < 1$ AU} perish.  These
numbers are much higher than Bortle's because of improved
quality of the 2000--2017~observations.  Besides the controversial
comet C/2016~U1 there are only two survivors with small perihelion
distances --- C/2006~P1 (McNaught) and C/2011~L4 (PanSTARRS)
--- both spectacular and dust-rich comets (Section~6 and Table~6).
The correlation with the absolute
brightness is striking as well:\ only 14~percent of the
Oort Cloud comets with \mbox{$H_0 < 8$} perish but fully
86~percent with \mbox{$H_0 > 10$} do so.

\begin{figure}
\vspace{-2.77cm}
\hspace{0.88cm}
\centerline{
\scalebox{0.715}{
\includegraphics{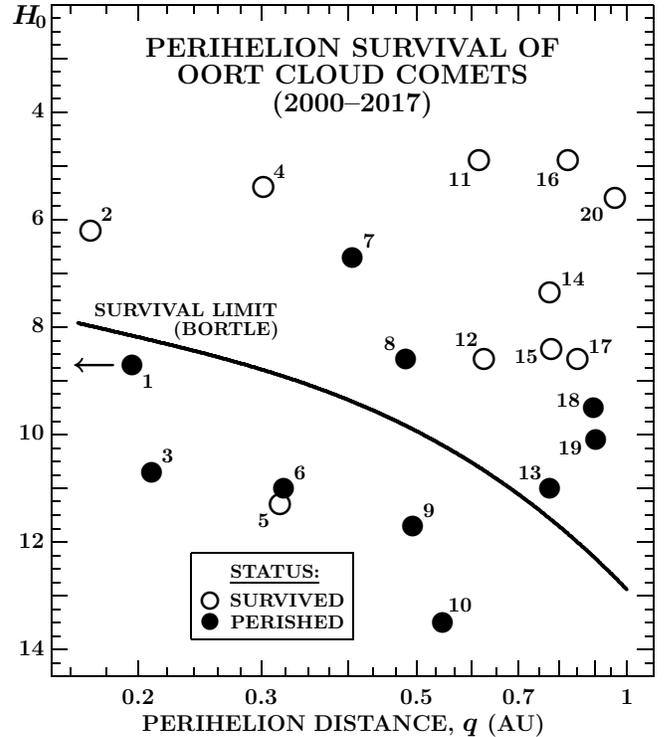}}}
\vspace{-9cm}
\caption{Oort Cloud comets discovered in 2000--2017 in
the plot of the absolute magnitude $H_0$ against perihelion
distance $q$.  Open circles are the surviving objects, solid
circles those perishing near perihelion.  The comets are
identified by their serial numbers from Table~2.  Comets
above the curve were predicted by Bortle (1991) to survive,
below it to perish.  The five perishing comets above the
curve (Nos.~7, 8, 13, 18, and 19) show that the propensity
of Oort Cloud comets for disintegration is stronger than
expected.  Only one comet, No.~5, survived while predicted
to perish.{\vspace{0.6cm}}}
\end{figure}

No less astounding are the results of comparison of
the examined comets with Bortle's formula; only one comet
(the controversial C/2016~U1) survives in the region below
the survival limit in Figure~1 (in fair agreement with Bortle's
expectation), but fully five perish --- 36~percent among the
comets expected to survive!  Hence, {\small \bf Bortle's formula
considerably underestimates the number of Oort Cloud comets that perish}. 

\section{Comets With Orbital Periods of\\2000--1\,000\,000
Years}
The procedure employed for the Oort Cloud comets was
next applied to the comets of the other two categories
defined at the end of Section~2, although no detailed
description is provided of observational evidence on
the status of the individual objects. It should be   
noted that for some of these comets it was more difficult
than for the Oort Cloud comets to decide whether they
survived or perished, primarily because of more limited
data available.  These complications notwithstanding,
the validity of the main conclusions below is unaffected.

\begin{table}[t]
\vspace{-4.2cm}
\hspace{4.3cm}
\centerline{
\scalebox{1}{
\includegraphics{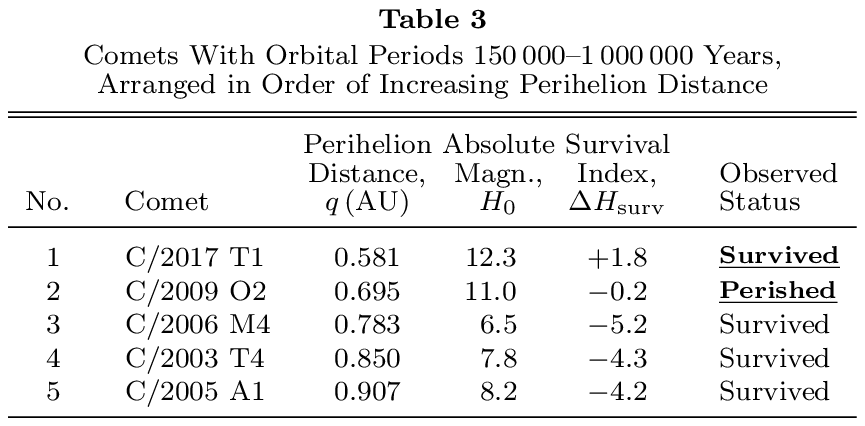}}}
\vspace{-20.75cm}
\end{table}

Consistent with dividing in Section~2 the non-Oort Cloud
comets into two categories, the results are presented
in Table~3 for the objects with the orbital periods
longer than 150\,000~years and in Table~4 with the
orbital periods shorter than 50\,000~years; the comets,
for which it was possible to determine their survival
status with less than full confidence, are tagged with
a question mark and the entries are in the following
assigned a weight of $\frac{1}{2}$.

\begin{table}[b]
\vspace{-3.6cm} 
\hspace{4.3cm}
\centerline{
\scalebox{1}{
\includegraphics{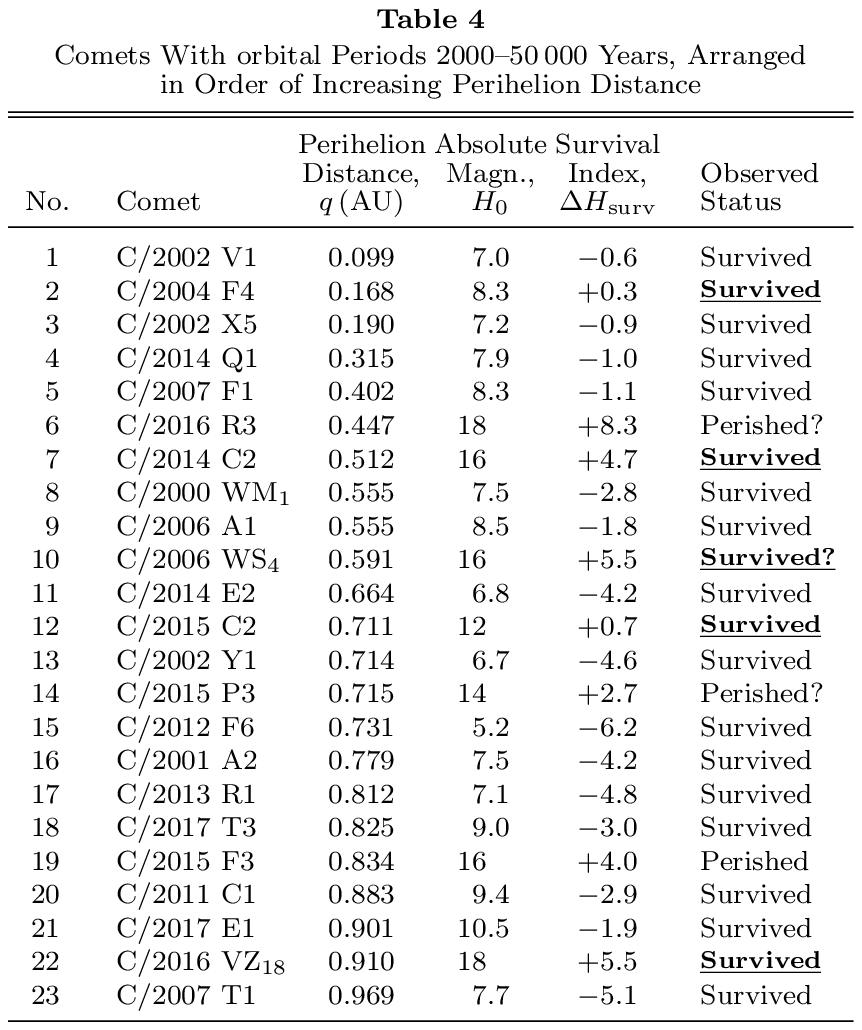}}}
\vspace{-15.3cm}
\end{table}

\begin{figure}[t]
\vspace{-0.98cm}
\hspace{0.86cm}
\centerline{
\scalebox{0.715}{
\includegraphics{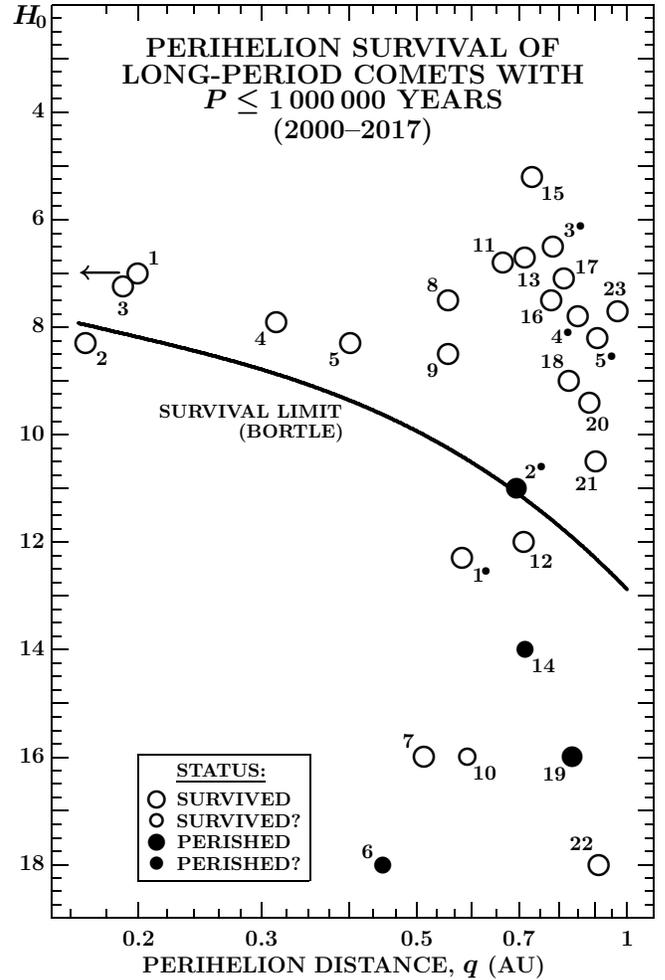}}}
\vspace{-7.57cm}
\caption{Long-period comets, not belonging to the Oort Cloud, discovered
in 2000--2017 in the plot of the absolute magnitude $H_0$ against perihelion
distance $q$.  Open circles of two different sizes represent objects observed
to have survived (larger symbols) or possibly survived perihelion, solid
circles of two different sizes show comets that perished (larger symbols)
or possibly perished near perihelion.  The individual objects are identified
by their serial numbers listed in Tables~3 and 4, the ones from Table~3 are
distinguished by a superscript dot following the digit.  It is
noted that, contrary to the Oort Cloud comets in Figure~1, an
overwhelming majority of these comets survive, showing no correlation
with the survival limit curve whatsoever.{\vspace{0.59cm}}}
\end{figure}

The two tables show that, unlike for the
Oort Cloud comets, only 12~percent of the comets in
Tables~3--4 perished, and only 5 and 15~percent of
those with perihelia below 0.6~AU and between 0.6~AU
and 1~AU, did so, respectively.  These numbers suffice
to show the enormous differences relative to the main
features of the Oort Cloud comets' statistics.  In
particular, as is clearly illustrated in Figure~2, the
fraction of perished comets is substantially depressed
and the dependence of their numbers on perihelion
distance tend to be, if anything, reversed, increasing
rather than decreasing with $q$.  On the other hand,
only about 25~percent of these comets defy Bortle's
rule, nearly all in the sense Bortle anticipated (i.e.,
with surviving comets below the threshold line).

To summarize the results of this section, the {\small
\bf non-Oort Cloud long-period~comets nearly all
survive}.  This category of comets appears
to have dominated the data set employed by Bortle (1991),
as comparison with the entries in Tables~3 and 4 suggests.

For the sake of completeness I present, in Table~5, the
12~comets in parabolic and poorly determined elliptic
orbits (Section~2).  Most of them were observed over
very short arcs of their orbits and they offer no
meaningful information on their perihelion survival
status.

\begin{table}
\vspace{-4.24cm}
\hspace{4.3cm}
\centerline{
\scalebox{1}{
\includegraphics{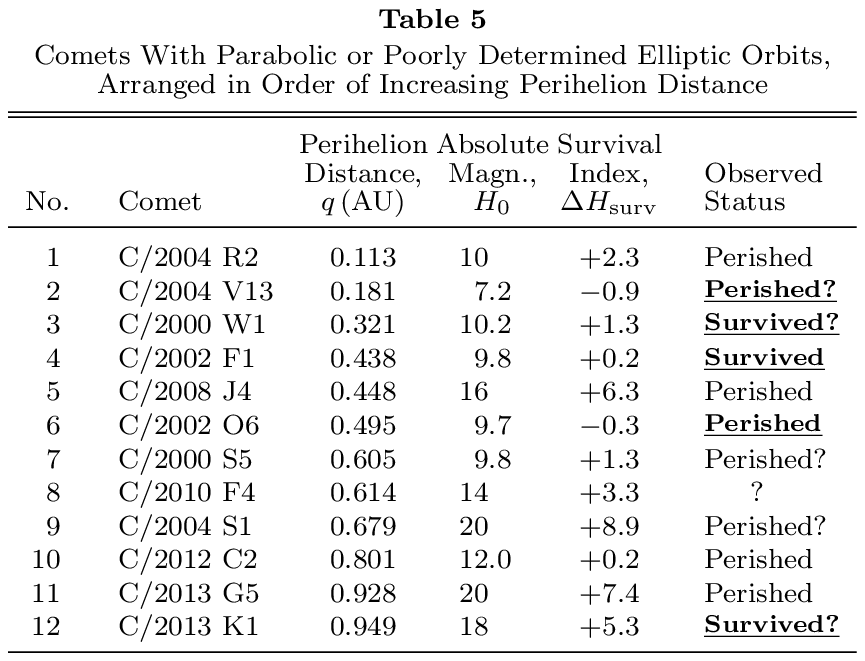}}}
\vspace{-18.37cm}
\end{table}

\section{Revision of Perihelion Survival Limit\\for
Oort Cloud Comets}
The results of the preceding sections indicate that the {\small
\bf nuclei of Oort Cloud comets differ dramatically in their
behavior near the Sun from the nuclei of other long-period
comets}.  The Oort Cloud comets account for virtually all instances
of disintegration reported by observers to begin as early as
several weeks before perihelion and manifested by the sudden
loss of the nuclear condensation usually, but not always, after
a brief flare-up.  There is also evidence that most, if not all,
of the perishing comets are poor dust producers.

The finding that it is {\it only\/} the Oort Cloud comets
that have this strong propensity for perishing near perihelion
provides a critical piece of evidence for the hypothesis
that an {\small \bf interstellar comet} of comparable
physical properties {\small \bf should have suffered a
similarly debilitating incident} shortly before perihelion,
{\small \bf with 1I/`Oumuamua emerging as its debris}.
An argument that a sizable fragment could never
survive the disintegration process is countered by
referring to Li \& Jewitt's (2015) study on one such
event, experienced by C/2010~X1; they were so struck by
the lack of vigor of the episode that they described the
disappearing comet --- in their paper's title --- as
``gone with a whimper, not a bang''.

Observations suggest that the process of disintegration of an
Oort Cloud comet is strongly perihelion-distance
dependent.  While a comprehensive modeling of the conditions
necessary for the nucleus' perihelion survival is beyond the 
scope of this paper, it appears that on its approach to
perihelion a {\small \bf comet perishes because it fails to
control effects of the steeply accelerating rate of increase
in the incident solar radiation flux}, ${\cal F}_{\mbox{\tiny
\boldmath $\!\odot$}}$ --- as if getting overheated.  I accept the
rate of flux increase as a proxy of the force that imperils the
comet's survival.  At time $t$ before the perihelion passage
$t_\pi$ (\mbox{$t < t_\pi$}), when the comet's heliocentric
distance is $r$, the rate of increase in the incident solar
flux is
\begin{equation}
\dot{\cal F}_{\mbox{\tiny \boldmath $\!\odot$}}(t) =
 \frac{d}{dr} \!\left( \!\frac{{\cal F}_0}{r^2} \!\right) \cdot
 \frac{dr}{dt},
\end{equation}
where ${\cal F}_0$ is the solar constant and $dr/dt$ can for Oort
Cloud comets be expressed with sufficient accuracy by a
parabolic approximation,
\begin{equation}
t = t_\pi - \frac{\sqrt{2}}{3k} (r\!+\!2q) \sqrt{r\!-\!q},
\end{equation}
with $k$ being the Gaussian gravitational constant.  The
rate of increase in the solar flux is then given by
\begin{equation}
\dot{\cal F}_{\mbox{\tiny \boldmath $\!\odot$}}(t) = \sqrt{8}
 \,k {\cal F}_0 \, r^{-\frac{7}{2}} \sqrt{1 \!-\! \frac{q}{r}},
\end{equation}
which implies \mbox{$\dot{\cal F}_{\mbox{\tiny \boldmath $\!\odot$}}(t_\pi)
 = 0$} and reaches a maximum of
\begin{equation}
(\dot{\cal F}_{\tiny \mbox{\boldmath $\!\odot$}}
  )_{\rm max} = \left( \:\!\!\frac{7}{8}\:\!\! \right)^{\!\!\frac{7}{2}}
 \!\!k {\cal F}_0 \, q^{-\frac{7}{2}} = C_0 \,q^{-\frac{7}{2}}
\end{equation}
at \mbox{$r_{\rm max} = \frac{8}{7}\,q$}, or at
\begin{equation}
t_{\rm max} = t_\pi - \frac{22}{21k} \sqrt{\frac{2}{7}}\, q^{\frac{3}{2}}
  = t_\pi - 32.55 \, q^{\frac{3}{2}},
\end{equation}
where $C_0$ is a constant; time is in days when $q$ is~in~AU. The
{\vspace{-0.06cm}}process of disintegration is completed before $t_{\rm
max}$, if it needs \mbox{$\dot{\cal F}_{\mbox{\tiny \boldmath $\odot$}} <
(\dot{\cal F}_{\mbox{\tiny \boldmath $\odot$}})_{\rm max}$}; it may continue
after $t_{\rm max}$,~even after perihelion passage, depending on the degree
of inertia.  For strongly hyperbolic orbits, see Appendix A.

Comets can cope with a high rate of increase in the incident solar flux
in a variety of ways.  Short-period comets with small perihelion
distance, such as 96P/Machholz or 2P/Encke, as well as long-period
comets with many perihelion passages in their past, such as C/2002~V1
or C/2004~F4, have long had the opportunity to
build up a surface mantle of sintered dust to improve strength.
Deprived of this option, Oort Cloud comets have to resort
to revving up their outgassing;
when almost all incident radiation is spent on
the sublimation of ices, the surface temperature is kept
nearly constant during approach to perihelion.  Observations
of Oort Cloud comets indicate that this response to increasing
solar radiation works well at heliocentric distances greater than
$\sim$1~AU.  At about this point in the orbit, however,
the objects' brightness is reported to grow at a progressively
slower rate and eventually to stall, an
apparent sign that the nucleus is running out of
near-surface supplies of ice.  The inevitable consequence of
this ice repository's exhaustion is the promptly increasing
surface temperature, with the undesirable implications, such
as surface material's thermal expansion, an upsurge in the
associated thermal stress, etc.
The deteriorating conditions cannot long be accommodated by
the devolatilized surface.  If the comet happens to reach
perihelion in the meantime, it might survive, otherwise it
surely disintegrates.\footnote{The disintegration (loss of
the nuclear condensation) is sometimes preceded by a brief flare-up,
apparently the evidence that a reservoir of ice deep in the
nucleus' interior was being tapped.}  The likelihood of survival
depends strongly on the perihelion distance, varies from object
to object, and is described by an accommodation limit $\cal A$.
When a comet of perihe\-lion distance $q$ is on the verge of
disintegration, one has
\begin{equation}
{\cal A} = (\dot{\cal F}_{\tiny \mbox{\boldmath $\!\odot$}})_{\rm max}
 = C_0 \, q^{-\frac{7}{2}}.
\end{equation}

\begin{table*}[t]
\vspace{-4.25cm}
\hspace{-0.5cm}
\centerline{
\scalebox{1}{
\includegraphics{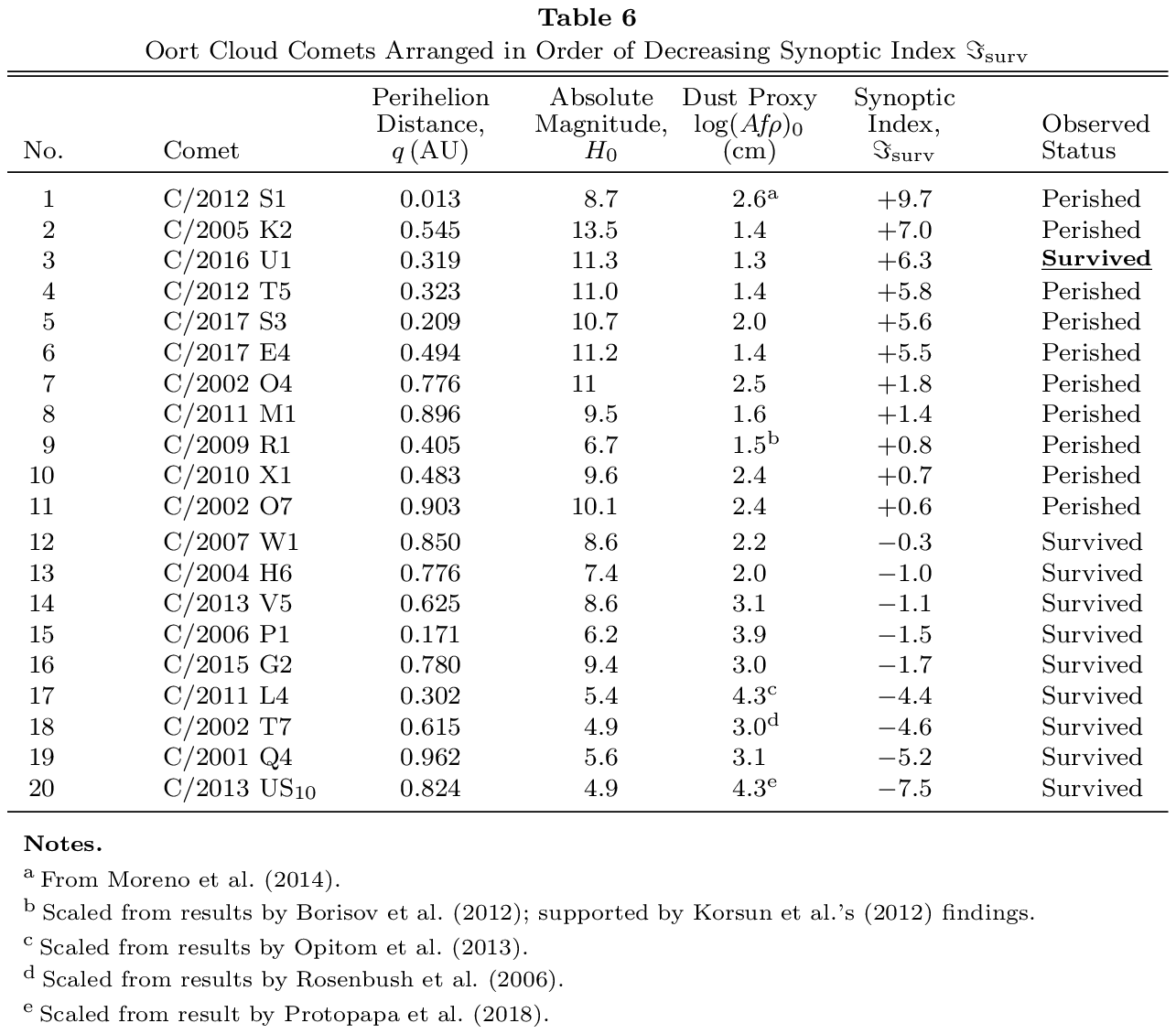}}}
\vspace{-13.47cm}
\end{table*}

Observations of Oort Cloud comets further indicate,
as already pointed out, that the perihelion survival
limit is{\nopagebreak} a function of the comet mass, $\cal M$:\
brighter (and presumably more massive) comets appear to
be able to tolerate higher rates of increase in the solar
flux than fainter (less massive) comets.  Accordingly,
\begin{equation}
{\cal A} \; \mbox{\raisebox{0.3ex}{\footnotesize $\propto$}} \, {\cal M} =
 c_0 10^{-\frac{3}{5} H_0},
\end{equation}
where the absolute magnitude $H_0$ is assumed to vary as
the nucleus' surface area and the nucleus' mass as a power
of $\frac{3}{2}$ of the surface area; \mbox{$c_0 = 2.5 \!
\times \! 10^{19}$\,g} according to Whipple (1975).  Comparing
relations~(6) and (7), one obtains a relationship between the
perihelion distance and the absolute magnitude that offers a
condition for the perihelion survival limit:
\begin{equation}
H_0 = c + \frac{35}{6} \log q, 
\end{equation}
where $c$ is a constant.  Comparison with the data points
in Figure~1 shows that this relation, like Bortle's
formula, fails to discriminate between a number of
surviving and perishing comets, implying that
the condition for the perihelion survival limit is
more complex.

It is the particulate dust in the atmosphere of an Oort Cloud
comet that makes the difference.
Dust-rich comets are more resistant than dust-poor comets
to damage from effects of a high rate of solar-flux increase.
As the rate of injection of dust from the nucleus increases,
the comet's atmosphere grows progressively less transparent,
until it eventually becomes optically thick, thereby
protecting the nucleus against further heating.

The accommodation limit $\cal A$ ought to show this aptitude of
{\vspace{-0.05cm}}dustier comets.  Since the rate of dust injection
$\dot{\cal M}_{\rm d}$ and its variation with heliocentric distance,
$r^{-m}$, are seldom known for Oort Cloud comets,
I replace $\dot{\cal M}_{\rm d}$ with a more readily available proxy
quantity \mbox{\Afro}, introduced by A'Hearn et al.\ (1984),
{\vspace{-0.05cm}}extensively examined by Fink \& Rubin (2012), and
related to $\dot{\cal M}_{\rm d}$ by
\begin{equation}
\dot{\cal M}_{\rm d}( r ) = (\dot{\cal M}_{\rm d})_0 \, r^{-m}
  = C_{\rm d} (\mbox{\Afro})_0 \, r^{-m},
\end{equation}
where $(\dot{\cal M}_{\rm d})_0$ and (\mbox{\Afro})$_0$ are, respectively,
the dust injection rate and the dust parameter at 1~AU from the Sun, and
$C_{\rm d}$ is a conversion coefficient that depends on the particle-size
distribution function.  With the quantity \mbox{\Afro} to be expressed in
cm, the improved model suggests for the accommodation limit $\cal A$
\begin{equation}
{\cal A} \; \mbox{\raisebox{0.3ex}{\footnotesize $\propto$}} \, {\cal M} \:\!\!
 \cdot \! (\dot{\cal M}_{\rm d})_0 \; \mbox{\raisebox{0.3ex}{\footnotesize
 $\propto$}} \, 10^{-\frac{3}{5} H_0} \!\cdot\! (\mbox{\Afro})_0.
\end{equation}
The condition (8) now changes to
\begin{equation}
H_0 = a + \frac{35}{6} \log q + \frac{5}{3} \log (\mbox{\Afro})_0.
\end{equation}

The last step in testing this procedure is the introduction of a {\small
\bf synoptic index for perihelion survival},~$\Im_{\rm surv}$, by
rearranging Eq.~(11) and choosing \mbox{$a = 5.7$}, so that, similarly
to $\Delta H_{\rm surv}$, \mbox{$\Im_{\rm surv} \!<\:\!\! 0$} when~a~comet is
predicted to survive perihelion essentially intact and \mbox{$\Im_{\rm
surv} \!>\! 0$}~when it is predicted to perish:
\begin{equation}
\Im_{\rm surv} = H_0 - 5.7 - \frac{35}{6} \log q - \frac{5}{3} \log
 (\mbox{\Afro})_0.
\end{equation}

\begin{figure}[t]
\vspace{-2.25cm}
\hspace{0.85cm}
\centerline{
\scalebox{0.69}{
\includegraphics{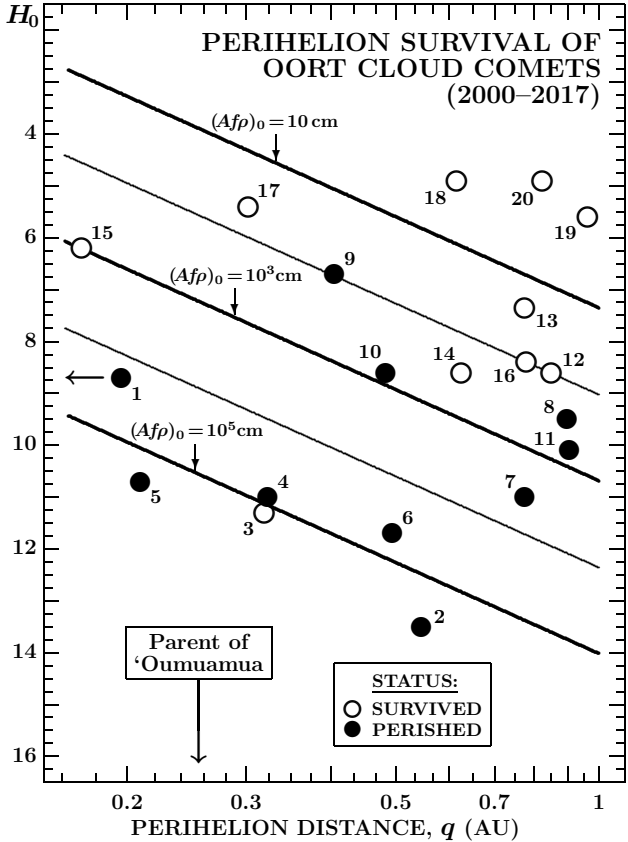}}}
\vspace{-7.3cm}
\caption{Oort Cloud comets discovered in 2000-2017 in the plot of the absolute
magnitude $H_0$ against perihelion distance $q$, with the lines of
a{\vspace{-0.04cm}} constant normalized dust parameter (\mbox{\Afro})$_0$
between 10~cm and 10$^5$\,cm.  The comets are identified by their serial
numbers from Table~6.  Only the controversial comet C/2016~U1 (No.~3)
does not fit the three dimensional relationship among $q$, $H_0$, and
(\mbox{\Afro})$_0$.  Also marked in the plot is the position of `Oumuamua's
parent.  See the caption to Figure~1 for more explanation.{\vspace{0.55cm}}}
\end{figure}

The 20 Oort Cloud comets from Table 2, rearranged by{\nopagebreak} decreasing index
$\Im_{\rm surv}$, are listed in Table~6.~Columns~2{\nopagebreak} to 4 are
copied from Table~2, whereas (\mbox{\Afro})$_0$ in column 5 has extensively been
researched in the literature.  For most entries in the table the starting value
was taken from the Spanish website maintained by J.~Castellano, E.~Reina, and
R.~Naves.\footnote{See {\tt http://astrosurf.com/cometas-obs}.}  For five
comets the starting data were taken from other sources, listed in the footnotes
to Table~6.  The value of \mbox{\Afro} at $\sim$1~AU from the Sun was used
when available, otherwise the value closest to 1~AU.  If a series of
data over a fairly wide interval of heliocentric distances showed that
\mbox{\Afro} was essentially constant then this value was identified with
(\mbox{\Afro})$_0$; if not, the tabulated value was reduced to 1~AU assuming
an $r^{-2}$ variation. The resulting value of (\mbox{\Afro})$_0$ was then
normalized to a zero phase angle employing either the dust-rich or
dust-poor version of the model developed by Marcus (2007).  It is this
value that is listed in column~5 of Table~6.  Column~6 presents the
index $\Im_{\rm surv}$ computed from Eq.~(12) and column~7, again copied
from Table~2, allows one to compare the prediction based on $\Im_{\rm
surv}$ with each comet's actual status.

\begin{figure}[t]
\vspace{-3.47cm}
\hspace{0.52cm}
\centerline{
\scalebox{0.66}{
\includegraphics{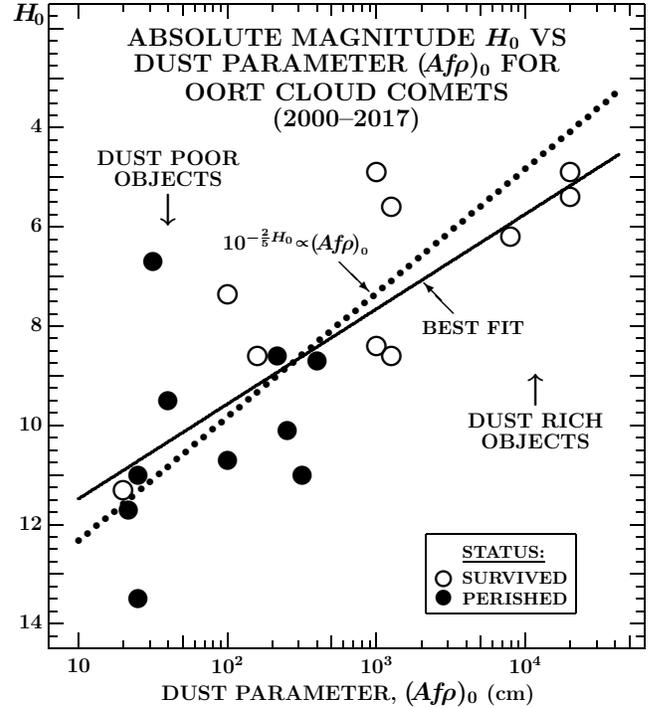}}}
\vspace{-7.02cm}
\caption{Plot of the absolute magnitude $H_0$ against the dust parameter
\mbox{(\Afro)$_0$} for the 20 Oort Cloud comets discovered betqeen 2000
and 2017.  The surviving and perishing objects are again represented
by different symbols.  The dotted line is a fit expected on the assumption
that both the absolute brightness and the dust production rate vary as
the surface area of the nucleus.  Striking is a concentration of the
perishing comets, mostly intrinsically faint and dust poor, to the lower
left, whereas the surviving comets, mostly  bright and dust rich, are to
the upper right.{\vspace{0.45cm}}}
\end{figure}

Except for the controversial object C/2016~U1, this new
index $\Im_{\rm surv}$ correctly discriminates between the
surviving and perishing comets.  One immediately notices a strong
concentration of the intrinsically faint and dust poor comets toward
the top of the table, while the bright and dust rich comets heavily
prevail near the bottom.

The introduction of \mbox{\Afro} allows one to represent the perihelion
survival limit (\mbox{$\Im_{\rm surv} = 0$}) in the plot of $H_0$ against
$\log q$ as a system of parallel straight lines shown in Figure~3.  An
object is predicted to survive perihelion when located above the line
with the respective~value~of (\mbox{\Afro})$_0$, but to perish when below
the line.~For~\mbox{example}, C/2005~K2 {\vspace{-0.06cm}}(No.\ 2)
and C/2017~S3 (No.\ 5) are predicted to perish even if their
(\mbox{\Afro})$_0$ was as high as 10$^5$\,cm, whereas C/2002~T7
(No.\ 18), C/2001~Q4 (No.\ 19), and C/2013~US$_{10}$ (No.\ 20)
are predicted to survive even if their (\mbox{\Afro})$_0$ was
as low as 10~cm.

Also marked in Figure~3 is the position of the parent comet of
`Oumuamua.  From the nondetection in the June 2017 PanSTARRS images
I estimated its peak absolute magnitude at \mbox{$H_0 \mbox{\gapeq} 18$}
(Sekanina 2019).  If similar to Oort Cloud comets in morphology and
other physical properties, the {\small \bf near-perihelion disintegration
of `Oumuamua's parent is absolutely inevitable}.

\begin{table*}
\vspace{-4.2cm}
\hspace{-0.52cm}
\centerline{
\scalebox{1}{
\includegraphics{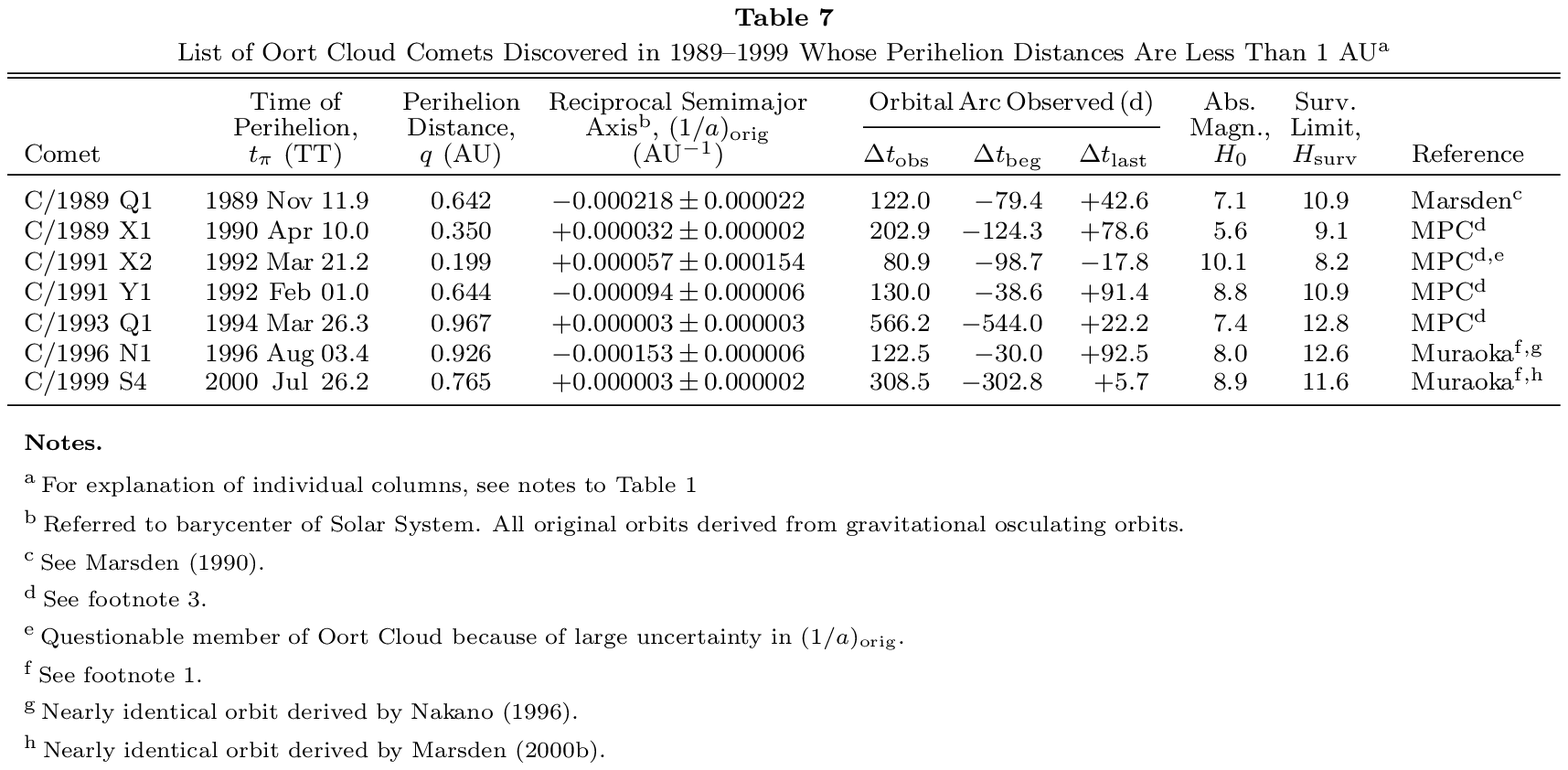}}}
\vspace{-16.37cm}
\end{table*}

The complete dominance of intrinsically bright and dust-rich
Oort Cloud comets among the perihelion survivors is likewise
illustrated in a plot of the absolute magnitude against the
dust parameter (\mbox{\Afro})$_0$, displayed in Figure~4.
Accepting an assumption that both the dust production rate
and the intrinsic brightness vary as the surface area of the
nucleus, one would expect that
\begin{equation}
H_0 = b - 2.5 \log (\mbox{\Afro})_0,
\end{equation}
where $b$ is a constant.  Fitting the 20 Oort Cloud comets from
2000--2017 to this relation yields \mbox{$b = 14.8$\,$\pm$\,1.8}.
On the other hand, the best least-squares fit is
\begin{eqnarray}
H_0 & = & 13.4 - 1.9 \log (\mbox{\Afro})_0. \\[-0.1cm]
    &   & \hspace{-0.1cm}\pm 1.1 \hspace{0.06cm} \pm \! 0.4 \nonumber
\end{eqnarray}
The modest difference between (13) and (14) is caused by scatter
among individual comets; for example, removing C/2009~R1 changes
the slope to $-$2.15\,$\pm$\,0.37.

\section{Examining Earlier Oort Cloud Comets}
There were reasons for limiting this investigation to the
period of time beginning in 2000.  Nakano's series of 
high-accuracy orbit determination with a full display
of positional residuals focused almost exclusively on
short-period comets before 2000; Kammerer started his
catalog of light curves with the comets of 1997; and
the collection of the parameter \mbox{\Afro} in the
Spanish website dates back --- with very few exceptions
--- to 1999.\footnote{On the other hand, a set of
85~comets by A'Hearn et al.\ (1995), incomplete as it
is, terminates in 1992.}

Nonetheless, since Bortle's set of long-period comets ended
with the 1988 objects, I considered it appropriate that
this study include a condensed report on the Oort Cloud
comets from the intervening period of 1989--1999.

The procedure that was employed to compile Table 1 was now
used to present a set of the Oort Cloud comets from this
11-year span with orbits of known quality.  The results
of this effort in Table~7 show that the number of such
dynamically new comets between 1989 and 1999 totals seven
at best, as one of them is burdened with an error so large
that its aphelion distance is entirely indeterminate, even
though the nominal value places the comet near the inner
boundary of the Oort Cloud.  The remaining six entries
are secure, the original orbits always determined from
gravitational solutions, despite the shorter orbital arcs
that had to be used (to avoid systematic trends in the
residuals), thus involving larger formal errors
than misleading results based on nongravitational
solutions would imply.  The light curves of the seven
comets and their parameters were published by Machholz
(1994, 1996) and by Shanklin (1997, 1998, 2001, 2009).  An
overlap of Shanklin's (2009) and Kammerer's (footnote~4)
light curves in 1999 shows they are in excellent agreement
with one another.

\mbox{The 1989--1999 collection, which delivers only margin-}
ally more than just one Oort Cloud comet per two years (compared to
more than one per year in the 2000--2017 period), is
incomplete, as several additional likely members
of the Oort Cloud are, judging from their physical
behavior, cataloged among the comets with
parabolic orbits --- the same problem that Bortle was
confronted with to a much greater extent in his
\mbox{1800--1988}~orbital data set.

Chronologically the first of the 1989--1999 Oort Cloud comets,
C/1989~Q1 (Okazaki-Levy-Rudenko; old designation \mbox{1989r =
1989~XIX}), was observed for astrometry over a post-perihelion
period of more than 40~days, visually for nearly 8~weeks.
Observations of \mbox{\Afro} were reported by A'Hearn et al.\ (1995)
and the original orbit was computed by Marsden (1990).  Even though the
post-perihelion fading was steeper than the preperihelion brightening
(common among Oort Cloud comets), the comet became more condensed as it
neared perihelion and retained this appearance until the end of
observations (Machholz 1996), showing no sign of disintegration
whatsoever.

The next comet in the table, C/1989 X1 (Austin; old designation
\mbox{1989c$_1$ = 1989~V}), was observed for about eleven weeks after
perihelion, despite its steeper post-perihelion fading, similar to
that of C/1989~Q1 (Machholz 1994).  Data on \mbox{\Afro} were reported by
Schleicher \& Osip (1990), by Osip et al.\ (1993), and by A'Hearn et
al.\ (1995).  There are no doubts that the comet survived perihelion.

\begin{table*}
\vspace{-4.2cm}
\hspace{-0.5cm}
\centerline{
\scalebox{1}{
\includegraphics{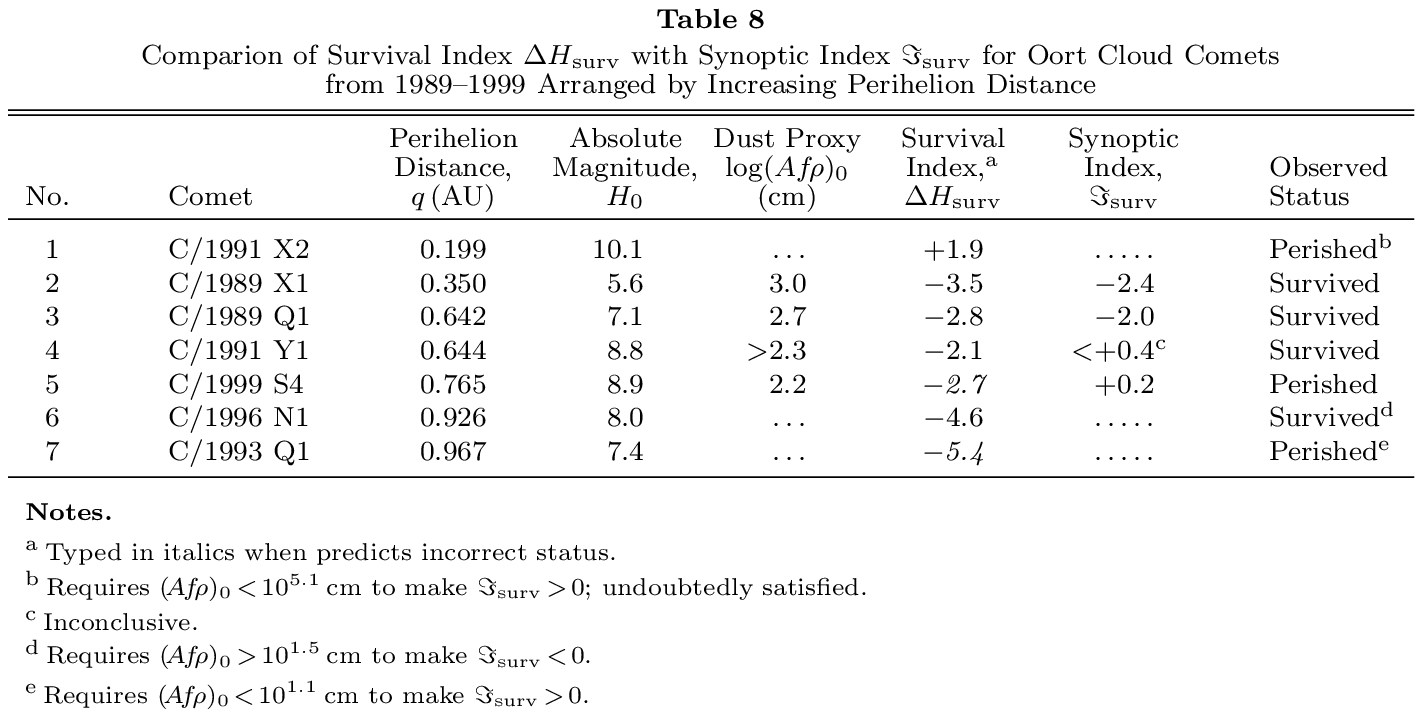}}}
\vspace{-17.47cm}
\end{table*}

A different story is C/1991 X2 (Mueller; old designation \mbox{1991h$_1$ =
1992 VIII}), whose astrometric and visual observations were terminated nearly
three weeks before perihelion because of the comet's rapidly decreasing
elongation.  No information on \mbox{\Afro} appears to be available.  Within
three weeks after perihelion, the comet was unsuccessfully searched for
in the infrared (Gehrz 1992) as well as at optical wavelengths (Hale 1992;
Kobayashi 1992; Seki 1992).  Shanklin (1997) concluded that the comet had
failed to survive.  The observed arc of the orbit is too short to determine
a set of high-quality elements (MPC; see footnote 3), so the comet's Oort
Cloud membership is uncertain, even though its near-perihelion disappearance
provides an argument to the contrary.  

Discovered only 10 days later, C/1991 Y1 (Zanotta-Brewington; old designation
\mbox{1991g$_1$ = 1992 III}) was a survivor, despite its asymmetric light
curve with a steeper post-perihelion fading (Shanklin 1997).  The only
\mbox{\Afro} data for this comet were obtained by Jorda et al.\ (1995),
unfortunately after perihelion.  The conversion to the normalized
preperihelion value to be used in the expression for the synoptic index is
burdened by large uncertainties, in part because at the time of Jorda et
al.'s observation the comet's brightness already subsided dramatically.
Under the circumstances, I only could try to estimate a lower limit to the
necessary correction.  The astrometry was obtained over a period of three
months after perihelion and the comet's survival is indisputable.

A peculiar case is C/1993 Q1 (Mueller; old designation \mbox{1993p = 1994 IX}),
which continued to brighten for two weeks past its perihelion near 1~AU.  The
last astrometric observation was made about one week later.  Only at that point
in the orbit did the comet's appearance suddenly changed; the disappearance of
the nucleus' condensation and dramatic fading were reported by Gilmore (1994)
on exposures taken 39--40~days after perihelion and by Camilleri (1994)
visually a week later, and confirmed by Scotti (1994) by images taken
another three or so weeks later.     

Comet C/1996 N1 (Brewington), discovered only one month before
perihelion, experienced an outburst of 1~mag in amplitude five
days after perihelion.  The event was not, however, followed by
the loss of nuclear condensation, perhaps in part because of
the perihelion distance that exceeded 0.9~AU.  The comet was
observed for astrometry for three months after perihelion and
the indications are that it survived.

The last member of this group of 1989--1999 comets, C/1999~S4
(LINEAR), was a major surprise.  I return to the dramatic changes
in the comet's appearance and the implications in Section~8; here
I only note that, in spite of its relatively large perihelion
distance, the comet did undergo a flare-up a few days before
perihelion, followed by the ominous disappearance of the nucleus'
condensation.  The difference compared to many previous (as well
as subsequent) similar occasions is that the events evolving
{\it after\/} the flare-up's termination were under scrutiny with the
help of the Hubble Space Telescope (HST) and the Very Large
Telescope (VLT), then the most powerful ground-based instrument
(Weaver et al.\ 2001).  The condensation's ``disappearance''
turned out under this in-depth view to consist of more than
a dozen minicomets that were being rapidly deactivated, the
foolproof evidence that the comet had perished.  The \mbox{\Afro}
data were measured and studied extensively by Farnham et al.\
(2001) and by Bonev et al.\ (2002); the data from a broad range
of heliocentric distances were averaged to derive an effective
value. 

The observed status of the 1989--1999 Oort Cloud comets, arranged
by increasing perihelion distance, is in Table~8 compared with
both indices, $\Delta H_{\rm surv}$ and $\Im_{\rm surv}$.  The
synoptic index could not be evaluated for three objects and only
constrained for a fourth one because of missing or incomplete
data.  The table shows that the index $\Delta H_{\rm surv}$ predicts
survival (typed in italics) for two perished comets, C/1993~Q1 and
C/1999~S4, a fraction of the set similar to that in the 2000--2017
collection.

\section{Morphology and Size of Debris Left Behind\\by Perished Oort
 Cloud Comets}
Reports on the appearance of most perished Oort Cloud comets in
their final stage of development are limited to acknowledging that
they either completely vanished or turned into a barely detectable
diffuse cloud with no evidence of a nucleus' condensation.  There
are only four instances with some information provided on what did
the process of ``disappearance'' mean in terms of the dimensions
of the surviving debris, with a potential of application to `Oumuamua:\
C/1999~S4, C/2010~X1, C/2012~S1, and C/2017~S3.

The most extensive amount of information was collected
for C/1999~S4, which was observed by the Hubble Space Telescope
(HST) even before it began to show the clear signs of disintegration
(Weaver et al.\ 2001).  Three weeks before perihelion the brightness
of the innermost coma (within 100~km of the nucleus) exhibited major
variations on a time scale of tens of hours, possibly associated
with the release of a small fragment.  The light curve reached
a broad maximum at about this time, but a gradual decline was
suddenly interrupted by a sharp flare-up, peaking three days
before perihelion, confirmed by an analysis of the dust tail
(Weaver et al.\ 2001) and coinciding with a peak in the
production of water (M\"{a}kinen et al.\ 2001).  Two days later
the coma became clearly cigar shaped (Kidger 2002), rapidly
developing into a long dust tail with a relatively sharp tip
devoid of any condensation in the ground-based observations.
Closeup imaging of this tip by the HST 10~days after perihelion
revealed a cluster of more than a dozen fragments, each resembling
a miniature comet with its own coma and tail.  Some 37~hours later
the head of the comet was imaged with the Very Large Telescope
(VLT) to show about 16~fragments.  A photometric and dynamical
investigation of the HST and VLT images showed that the fragments
were 50 to 120~meters in diameter (assuming a geometric albedo of
4~percent), at least some of them released long before perihelion,
and typically subjected to nongravitational accelerations,
probably outgassing driven, of about \mbox{$10 \times\! 10^{-8}$\,AU
day$^{-2}$} (Weaver et al.\ 2001), less than 50~percent of
the nongravitational acceleration affecting the orbital motion of
`Oumuamua.  Significantly, no fragments were detected in the VLT
images taken 12 and 17~days after perihelion, i.e., 15 and 20~days
after the terminal flare-up, fading by a factor of 2 to 10 in
72~hours.  An obvious conclusion is that the disintegration process
of C/1999~S4 proceeded very rapidly.  The comet's surviving dust
tail was detected by G.~J.~Garradd\footnote{See NASA website {\tt
https://apod.nasa.gov/apod/image/0009 /c99S4linear\_000821\_gg1.jpg}.}
with his 45-cm f/5.4 reflector two weeks after the last VLT detection
of the fragments.

The sublimation area of C/1999~S4,~determined from the
water production rates by M\"{a}kinen et al.\ (2001) based on their SWAN
observations over a period~from two to one month before perihelion, amounts
to 1.4~km$^2$, while the nucleus' projected area derived on the assumption of
a spherical nucleus from the nongravitational acceleration affecting the
{\vspace{-0.04cm}}comet's orbital motion (Marsden \& Williams 2008) is at
most 0.5~km$^2$.  This enormous disparity implies that even {\it before\/}
the disintegration event the nucleus may have been strongly nonspherical;
the assumption of a pancake-like shape leads to a long-to-short
dimension ratio of about 5.5, reminiscent of `Oumuamua's inferred
shape.  Alternatively, the nucleus may have been very ``fluffy'',
with the effective bulk density much lower than assumed, thereby
compromising the determination of the projected area from the mass
of the dynamical model, an issue that is also highly relevant to `Oumuamua.

With a perihelion distance substantially smaller than C/1999~S4,
comet C/2010~X1 experienced a flare-up more than three weeks before
perihelion (Li \& Jewitt 2015; Sekanina 2011) and almost
immediately began to lose the nucleus' condensation (Mattiazzo
\& McNaught 2011).  Li \& Jewitt (2015) illustrate the dramatic
change in the comet's appearance imaged with the same telescope
five weeks apart in their Figure~4.  The comet was too
close to the Sun in the sky for ground-based observing from
shortly before perihelion until one
month after perihelion.  Li \& Jewitt used the 360-cm CFHT reflector
to image the comet's expected position 40~days after perihelion,
but without success.  They estimated that any potentially surviving
fragments could not be larger than 80~meters across.  Much larger
upper limit on the size of nuclear fragments was reported by Kidger
et al.\ (2016) from their negative observations with the Herschel
Space Observatory.

Comet C/2012 S1 was an extreme case because of the exceptionally small
perihelion distance of 2.7~solar radii.  Rather than providing a summary
of the various investigations published on this comet, I paraphrase
Sekanina \& Kracht's (2014) proposed timeline of events, based on such
studies.  This time interval began 16~days before perihelion at a
heliocentric distance of 0.7~AU with a minor increase in the water
production that was followed by a major outburst, when the production
went up by a factor of 16.  This surge suggested that a new source
of water was tapped, requiring fragmentation of the nucleus.  A few
days later there was another production jump, by a factor of three,
implying further fragmentation.  Then about three days before perihelion
came a dramatic drop in the gas production, suggesting that the comet's
reservoir of ice was practically exhausted.  The comet {\it de facto\/}
ceased to exist already at this time, except that the sublimation of
sodium was still increasing and the process of cascading fragmentation
continuing. All post-perihelion images taken by the coronagraphs on board
the SOHO and STEREO spacecraft consistently show that dust was ejected
from nuclear fragments until 3.5~hours before perihelion, when at
5~solar radii the process had nothing to feed itself on any more ---
the ultimate termination of activity.  While Knight \& Battams (2014)
estimated ``any remaining active nucleus [at] $<10$~meters in radius'',
Sekanina \& Kracht's (2014) scenario led them to the conclusions
that the largest surviving {\it inert fragments\/} of the nucleus
were at best pebble sized and that {\it no active\/} nucleus survived,
a notion supported by Curdt et al.'s (2014) failure to find any
Lyman-alpha emission less than one hour before perihelion. 

Comet C/2017 S3 underwent two preperihelion outbursts; the first began
nearly seven weeks before perihelion at a heliocentric distance of
1.25~AU, the second two weeks later at 0.96~AU (Sekanina \& Kracht
2018).  The nucleus appears to have survived the first outburst with
only minor damage, releasing --- as later determined (see below) ---
a companion undetected in a condensation of simultaneously dust ejecta
before the onset of the second outburst.  In the course of this event
the nucleus was completely shattered into a massive cloud of rapidly
expanding dust cloud, which obliterated the ejecta from the first
outburst over a period of two weeks.  As the ground-based observations
were about to terminate, the measured astrometric positions ceased to
fit the debris from the second outburst and, instead, turned out to
be consistent with the location of the companion released in the
first outburst, implying its radial nongravitational acceleration
of \mbox{$16.9 \times \! 10^{-8}$AU day$^{-2}$}, about 70~percent of
`Oumuamua's acceleration.  In the absence of activity, this is an effect
of solar radiation pressure, suggesting that the companion had a very
high surface-area-to-mass ratio and represented an extremely fluffy
aggregate of loosely-bound dust grains, a conclusion supported by the
result derived directly from a nongravitational orbital solution linking
all astrometric observations made prior to the second outburst.     

These four examples show that for different perishing comets the process
of disintegration does not pass through identical stages.  None of them can
serve as an analog for the parent comet of `Oumuamua, a dwarf object with
the absolute magnitude $\ge$18, because they all were much too bright (and
massive), as documented in Table~2.  Nonetheless, the apparent fragment of
C/2017~S3 released during the first outburst and astrometrically measured
for several days about one month later appears to be an object whose role
in the disintegration process may have been a little like that of `Oumuamua,
even though information on it is extremely limited.  The data on C/1999~S4
suggest that fragments get rapidly depleted of ice; whether they survived
as large boulders or were crumbled into pebbles and/or  dust is unclear.

There are some dwarfs in Table~4 among the long-period comets in non-Oort
Cloud orbits, but these are of course irrelevant.  No useful information
comes from cursory inspection of Table~5 containing a few dwarf objects in
poorly known orbits, all of perihelion distance larger than is of interest,
and appearing, with the exception of C/2013~K1, preperihelion.

\section{Constraints on Shape, Dimensions, and Mass of `Oumuamua}
I submit that the Spitzer Space Telescope's failure to detect `Oumuamua
(Trilling et al.\ 2018) provides~{\small \bf strong evidence for preferring
pancake over cigar shape}.~This argument is supported by simply
comparing the two models.  Let the maximum projected area derived from
the peak absolute brightness during the tumbling be $X_{\rm max}$.  Let
the light curve's amplitude be $2.5 \log \gamma$, where \mbox{$\gamma
\!>\! 1$} is the ratio of the maximum-to-minimum projected area.  The
long and short diameters of `Oumuamua in the case of cigar shape equal,
respectively,
\begin{eqnarray}
(D_{\rm max})_{\rm cig} & = & \sqrt{\frac{4 \gamma}{\pi} X_{\rm max}}
  \nonumber \\
(D_{\rm min})_{\rm cig} & = & \sqrt{\frac{4}{\gamma \pi} X_{\rm max}}
  = \frac{(D_{\rm max})_{\rm cig}}{\gamma}.
\end{eqnarray}
The maximum and minimum diameters in the case of pancake shape amount to
\begin{eqnarray}
(D_{\rm max})_{\rm pan} & = & \sqrt{\frac{4}{\pi} X_{\rm max}} \nonumber \\
(D_{\rm min})_{\rm pan} & = & \frac{1}{\gamma} \sqrt{\frac{4}{\pi} X_{\rm
  max}} = \frac{(D_{\rm max})_{\rm pan}}{\gamma} .
\end{eqnarray}
To determine $X_{\rm max}$ I adopt a peak $R$ absolute magnitude of 21.7
by Drahus et al.\ (2018), more conservative than Jewitt et al.\ (2017)
value, which~is~0.2~mag brighter.  Since Trilling et al.\ (2018) present
their results as a function of a visual albedo, I converted the
$R$~magnitude to $V$ with Jewitt et al.'s color index \mbox{$V \!-\! R =
+0.45$}; this gives \mbox{$X_{\rm max} = 0.002/p_V$ km$^2$}, where $p_V$
is the visual geometric albedo.  Trilling et al.\ provide upper limits of
the {\vspace{-0.09cm}}dimensions for three different albedos, varying of
course as $p_V^{-1/2}$ and being equivalent.  Selecting{\vspace{-0.04cm}}
 \mbox{$p_V \!=\! 0.1$}, one of the three options, I determine
\mbox{$X_{\rm max} = 0.02$ km$^2$}, compared to Jewitt et al.'s
(2017) 0.025~km$^2$.  I accept Trilling et al.'s $\gamma$ value
of 6 to make the comparison fully compatible.  Equations~(15) and
(16) result in \mbox{$(D_{\rm max})_{\rm cig} = 391$ m} and
\mbox{$(D_{\rm min})_{\rm cig} = 65$ m} for cigar shape, but
\mbox{$(D_{\rm max})_{\rm pan} = 160$ m} and \mbox{$(D_{\rm
min})_{\rm pan} = 27$ m} for pancake shape.  {\small \bf Trilling
et al.'s 3{\boldmath $\sigma$} upper limits} are 341~m for the
maximum diameter and 57~m for the minimum diameter, thus being
{\small \bf compatible only with the pancake-like model}.

The cigar-like and pancake-like configurations differ from one another in the
object's volume, $Y$, as well.  Approximations by the prolate and oblate
spheroids give, respectively,
\begin{eqnarray}
Y_{\rm cig} & = & \frac{\pi}{6} \frac{(D_{\rm max})_{\rm cig}^3}{\gamma^2}
  = \frac{4}{3 \sqrt{\gamma \pi}} X_{\rm max}^{\frac{3}{2}} \nonumber \\
Y_{\rm pan} & = & \frac{\pi}{6} \frac{(D_{\rm max})_{\rm pan}^3}{\gamma}    
  = \frac{4}{3 \gamma \sqrt{\pi}} X_{\rm max}^{\frac{3}{2}},
\end{eqnarray}
showing that the volume of the pancake-like model is $\sqrt{\gamma}$ times
smaller.  A pancake-like configuration is also preferred because of perceived
better dynamical stability for an extremely fluffy object.

Looking at the issue from the standpoint of the hypothesis that `Oumuamua
is a {\it highly irregularly shaped fragment\/} of an interstellar comet,
Equations~(15) through (17) indicate that the {\small \bf pancake-like model
is a positively better approximation} to the object's actual shape than the
cigar-like model.

The next point is the problem of extremely high porosity needed for
`Oumuamua as an aggregate of submicron-sized grains of dust to succeed.
I address two basic constraints that concern the object's mass:\ (i)~it
should satisfy the projected-area-to-mass ratio required by the detected
nongravitational acceleration in its orbital motion and (ii)~it should
{\it by orders of magnitude\/} exceed the absolute lower limit implied
by a cloud of unbound submicron-sized grains of equal projected area. 

\begin{figure*}[t]
\vspace{-8.9cm}
\hspace{-0.1cm}
\centerline{
\scalebox{0.9}{
\includegraphics{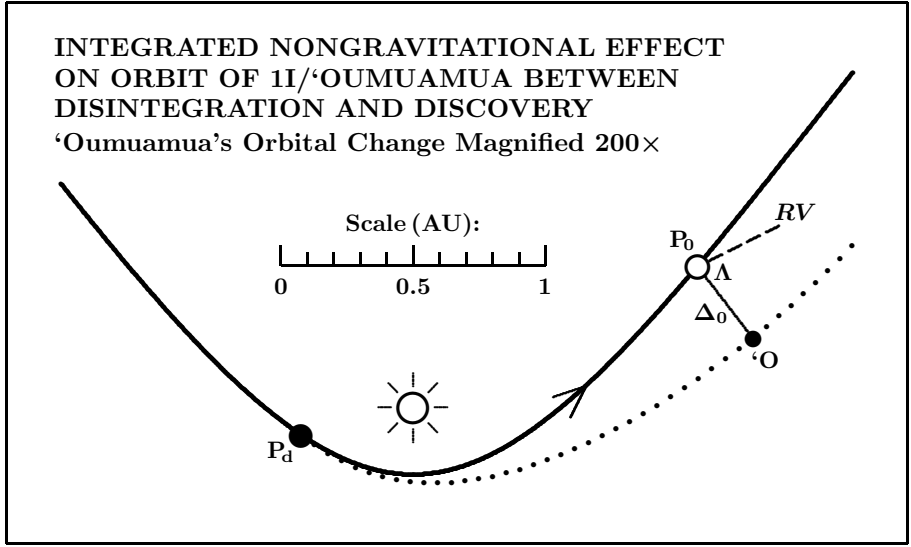}}}
\vspace{-10.6cm}
\caption{Integration of the orbital changes triggered by the nongravitational
acceleration of 1I/`Oumuamua between the time of disintegration of the parent
comet and the time of `Oumuamua's discovery.  The solid curve is the orbit of
the parent comet up to the point of its disintegration, P$_{\rm d}$, assumed
here to have occurred 10~days before perihelion.  Within a fraction of its
width, the solid curve is the orbit of `Oumuamua from the point P$_{\rm d}$
on.  The arrow indicates the direction of orbital motion.  The dotted curve
is the orbit of `Oumuamua after the integrated effect of its nongravitational
acceleration was magnified by a factor of 200.  At the time of `Oumuamua's
discovery, the extrapolated position of the parent is marked by P$_0$, the
prolonged radius vector, shown as a dashed line, by {\it RV\/}.  After
magnification of the orbital changes of `Oumuamua, its position at the time of
discovery is marked as `O, its distance from P$_0$ as $\Delta_0$, and the lag angle
of its position vector behind the radius vector as $\Lambda$.{\vspace{0.58cm}}}
\end{figure*}

The ratio of the geometric projected area $X$ to the mass ${\cal M}$ of
an object subjected to solar radiation pressure is given by
\begin{equation}
\frac{Q_{\rm pr}X}{\cal M} = \frac{4 \pi c G {\cal M}_{\mbox{\tiny \boldmath
 $\odot$}}}{{\cal L}_{\mbox{\tiny \boldmath $\odot$}}} \beta,
\end{equation}
where $c$ is the speed of light, $G$ is the universal gravitational
constant, ${\cal M}_{\mbox{\tiny \boldmath $\odot$}}$ and ${\cal L}_{\mbox{\tiny
\boldmath $\odot$}}$ are the mass and luminosity of the Sun, and
$\beta$ is a dimensionless, heliocentric-distance independent quantity
that measures an acceleration ratio of solar radiation pressure to solar
gravitational attraction.  Because of the scattering properties of dust
grains, their cross sectional area for radiation pressure differs
generally from their geometric cross sectional area.  To account for
this difference, Equation~(18) contains a dimensionless quantity
$Q_{\rm pr}$, the efficiency factor for radiation pressure that converts
the latter to the former, but is generally close to unity.\footnote{For
effects of radiation pressure on fluffy porous dust aggregates see,
e.g., Kimura \& Mann (1999), Tazaki \& Nomura (2014).}  With the
standard values of the object independent quantities and
\mbox{$\beta = 0.00083$} for `Oumuamua, one has{\vspace{-0.04cm}}
 \mbox{${\cal M} = 0.092 Q_{\rm pr} X$}, where $X$ is in~cm$^2$
and ${\cal M}$ in g.  Since the projected area varies quasi-periodically
with time while the nongravitational acceleration is expressed as an
averaged effect, it is likewise necessary to average the projected
area exposed to sunlight.  The method of averaging, which shows
the result to depend on the axial ratio, is described
in Appendix~B; here I note that for the range of observed axial
ratios the average projected area is 0.52 the peak area of 0.02~km$^2$.
Taking \mbox{$Q_{\rm pr} \simeq 1$} and rounding the result off, the
predicted mass of `Oumuamua comes out to be
\begin{equation}
{\cal M} \simeq 1 \!\times\! 10^7 \, {\rm g},
\end{equation}
identical to the mass estimated by Sekanina \& Kracht (2018).  With the
above dimensions this mass indicates that `Oumuamua's bulk density is
$\sim$0.00003~g~cm$^{-3}$.

Although only loosely interconnected, submicron-sized particles in an aggregate
should attenuate most incident sunlight, which requires that Oumuamua be orders
of magnitude more massive than an optically thin cloud of submicron-sized
grains of {\it equal\/} projected area.  Compliance with this categorical
condition is tested by comparing the mass of such a dust cloud with the above
mass estimate of the object.  Assuming that the aggregate consists of spherical
grains 0.2~$\mu$m in diameter, the projected area of a single grain is
\mbox{$X_{\rm gr} =  \pi \times\! 10^{-10}$\,cm$^2$}.  The number of grains
needed to equal the peak projected area of `Oumuamua is{\vspace{-0.05cm}}
\mbox{0.02~km$^2/X_{\rm gr} = 6.4 \times\! 10^{17}$}.  At an expected density
of 3~g~cm$^{-3}$, their total mass is 8000~g, more than a factor of $10^3$
{\it lower\/} than `Oumuamua's estimated mass.  On the average, only
0.08~percent of the projected surface of submicron-sized grains in the
fluffy aggregate is exposed to sunlight, a very small fraction.

\section{Integrated Orbital Changes Triggered by\\`Oumuamua's
Nongravitational Acceleration}
In the framework of the investigation of `Oumuamua as an interstellar
comet's fragment subjected to a nongravitational acceleration, one
issue of particular interest is the orbital effect integrated over a
period of time from the parent's disintegration to `Oumuamua's discovery.
An example is displayed in Figure 5 for the parent's disintegration time
of 10~days before perihelion.  The effect of an integrated nongravitational
acceleration is magnified by a factor of 200 for clarity.  The solid curve
is the orbit of the parent up to the point P$_{\rm d}$ and of `Oumuamua
from that point on (which coincides within the width of the drawn curve 
with the parent's extrapolated orbit), while the dotted curve is
`Oumuamua's orbit after the application of the magnification factor.
As seen, `Oumuamua moved in a slightly
larger orbit and was at the time of discovery lagging a little behind the
position it would occupy if there were no extra acceleration.  The radial
nongravitational acceleration of 0.000830\,$\pm$\,0.000027 the Sun's
gravitational acceleration, which was derived by Micheli et al.\ (2018),
has on `Oumuamua the same effect as if it were orbiting in a gravitational
field of the Sun whose mass was reduced to 0.999170\,$\pm$\,0.000027 its
actual mass.  The often noted ``velocity boost'' that `Oumuamua allegedly
received from the nongravitational acceleration is misleading.  The object's
location in the orbit relative to the unperturbed motion is time dependent
and a function of the point of separation from the parent.

\begin{table}[t]
\vspace{-4.18cm}
\hspace{4.22cm}
\centerline{
\scalebox{1}{
\includegraphics{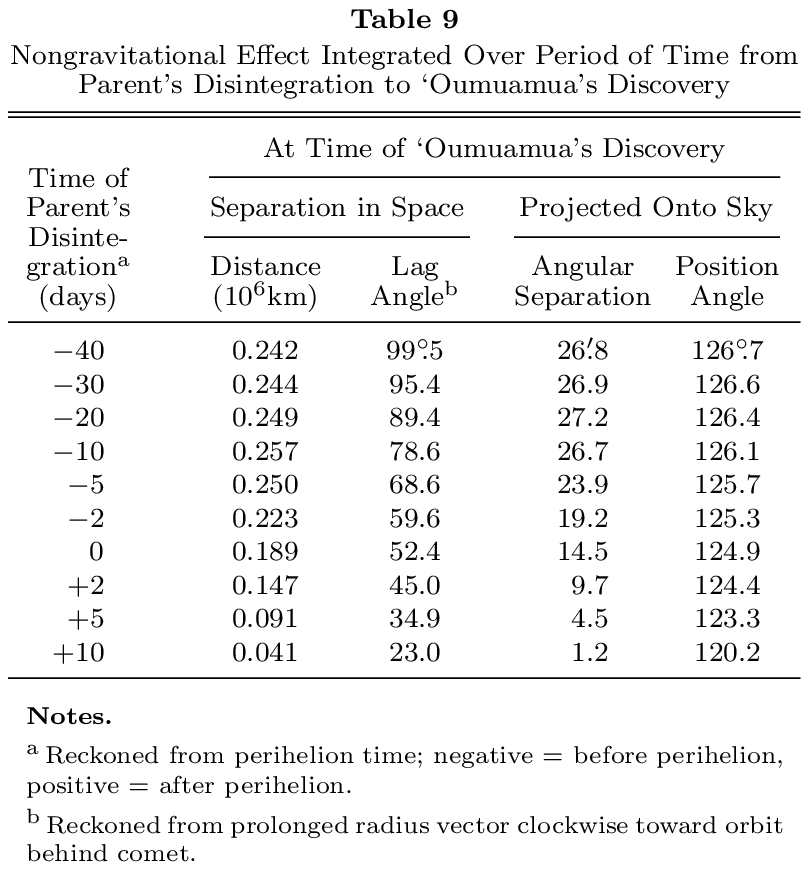}}}
\vspace{-16.2cm}
\end{table}

Since the disintegration time of the parent comet~is~not known, I assume for
it a number of times between 40~days before perihelion and 10~days after
perihelion.  `Oumuamua is presumed to have acquired no separation velocity;
the integrated effect is presented in Table~9 both as (1)~its separation
distance from the unperturbed position of the parent in the orbital
plane and the lag angle reckoned from the prolonged radius vector
toward the orbit behind the parent, as shown in Figure~5; and (2)~an angular
separation distance and the position angle in projection onto the plane of the sky.

Table 9 illustrates the enormity of the effect of the nongravitational
acceleration accumulated over only several weeks, with the separation distance
growing to an astonishing quarter of a million kilometers with a large lag angle.
This effect is of the same nature as the well-known striking post-perihelion
broadening of dust tails of comets with small perihelion distance, a
consequence of the law of conservation of orbital angular momentum.  In
practical terms it means that individual pieces of the debris that the
parent's disintegration resulted in were by the time of `Oumuamua's
discovery scattered over a huge volume of space, with a negligibly low
spatial density.  As obviously the largest of these fragments (subjected
to the lowest acceleration), `Oumuamua was the only one detected.  In
terms of the angular separation in the sky, the effect amounts to nearly
one half degree.  Again, other fragments, if there are any, could easily be
scattered over many degrees from the unperturbed position and would be missed
except perhaps by large wide-field instruments.  The peculiar lack of variation
in the position angle is caused by the Earth being located only 1$^\circ\!$.7
below the orbital plane as viewed from `Oumuamua.

\section{Final Comments and Conclusions}
The post-perihelion discovery of `Oumuamua has left open the question of its
relationship to the interstellar object that had entered the inner regions
of the Solar System:\ Was `Oumuamua identical to this object or did the
object experience shortly before or at perihelion a disintegration event
whose product `Oumuamua has become --- a large piece of debris of the
parent?  The answer is of course a matter of conjecture that
depends on the starting postulates on which the argument is formulated.
The fundamental postulate of this study is that the {\small
\bf object was an active interstellar comet}~whose~morphology~and
related physical properties were {\small \bf similar~to}~those~of the
{\small \bf Oort Cloud comets}, a plausible hypothesis given that both
were exposed to the same kind of environment over an extended period of
time.  In addition, in order to comply with the nondetection in the
PanSTARRS's June 2017 images, the comet's activity should have been very
low.  The object is therefore classified as a {\small \bf dwarf comet},
with the synoptic index exceeding +16.  Under these
conditions it is absolutely impossible to escape the conclusion that
{\small \bf `Oumuamua's parent inevitably perished near perihelion},
most probably shortly before perihelion, and that {\small \bf `Oumuamua
is a major fragment of the original comet}, possibly the only surviving
piece of debris of substantial mass.

`Oumuamua is often described as unlike any other cosmic object ever detected.
The present results question whether any known Oort Cloud comet is intrinsically
as faint or fainter.  The absolute magnitude of the faintest one among those
listed in this paper, C/2005~K2, is 13.5 (Tables~1, 2, and 6), compared to an
estimated absolute magnitude of $\ge$18 for `Oumuamua (Sekanina 2019).
For this reason, evidence on the debris of the best examined
perishing Oort Cloud comets (Section~8) is not~quite relevant to
`Oumuamua or its parent comet, except for obvious features, such as
the absence of outgassing by an ice depleted fragment.  The strongest
evidence dictating the extremely low bulk density and super fluffy
nature of `Oumuamua (Sekanina 2019) is of course the {\small \bf
detected nongravitational acceleration} (Micheli et al.\ 2018), which
could hardly be {\small \bf driven by} anything else than {\small \bf
solar radiation pressure}, given the nonexistence of activity.  As
noted in Section~8, a similar kind of object appears to have been briefly
detected in C/2017~S3, but only little relevant information is available.

The enormous amplitude of `Oumuamua's light curve is overwhelmingly
interpreted as shape driven, although a contribution from albedo variations
over the surface~is sometimes invoked (e.g., Jewitt et al.\ 2017).  The
clear preference for cigar shape over pancake shape in the literature may be
influenced by the fact that the several nuclei (of the short-period comets)
of known figure resemble a prolate rather than oblate spheroid.  However,
the critical {\small \bf nondetection of `Oumuamua by the Spitzer Space
Telescope implies compatibility with optical data for pancake shape but
not for cigar shape}.

The extremely high porosity is perhaps unique for a comet-like object,
but the estimated mass of `Oumuamua is still more than three orders of
magnitude higher than the mass of a cloud of unbound submicron-sized particles of
the same total projected area.  Accordingly, the implied fluffy structure
does significantly attenuate sunlight not only because of the low albedo.
It is expected that `Oumuamua's dimensions do not exceed the ones determined
from optical observations.

The magnitude of the nongavitational acceleration, equivalent to
\mbox{$83 \times \! 10^{-5}$} the solar gravitational attraction
is by no means minuscule.  Its integrated effect over the period of
time between the parent comet's presumed disintegration and `Oumuamua's 
discovery equals, depending on the former's time, up to more than
250\,000~km and nearly 0$^\circ\!$.5 in projection onto the plane
of the sky.  Whereas the spatial separation was of course increasing
with time, the projected separation was decreasing because of
`Oumuamua's rapidly increasing distance from the Earth.  A lesson
learned from this exercise is that~{\small \bf any potential lesser
fragments should have been scattered over a large area of the sky}
at the discovery time, when they were at their brightest for
terrestrial observers.

The breakup event of the parent comet near perihelion has, besides the
potentially overwhelming effect on `Oumuamua's morphology,
other implications.  One concerns the question of the delicate fluffy
structure's survival over a long period of time.  If this property were
intrinsic to the object before its arrival to the Solar System, one
would have to demonstrate that it was able to survive interstellar
travel.  In the hypothesis proposed in this paper, the structure's
lifetime on the order of 200~days would safely warrant its survival
over a period ending with the object's last observation; in fact,
moderate  instability and/or fragmentation may not have been
observationally recognized during this time span.  The other implication is
the impact on investigations aimed at establishing the stellar system
from which `Oumuamua had arrived.  Given that the time of the
disintegration event is unknown and that, in addition, the orbital
motion of the parent comet may have been affected by an outgassing-driven
nongravitational acceleration of unknown magnitude, the reliable
determination of the parent's initial incoming velocity in interstellar
space is unfortunately compromised.

We never have had a chance to inspect --- and~may~not have in the
foreseeable future --- a closeup image~of~the nucleus of an Oort Cloud
comet, and details of~its~nature remain unknown.  Comparison with
other long-period comets near the Sun suggests that {\small \bf Oort
Cloud comets are very different, showing the strong propensity for
disintegrating near perihelion, especially when close to the Sun}.
Their nuclei appear to {\small \bf experience~\mbox{major} problems in
tolerating the environment of relentless solar heating}, ever more
so as heliocentric distance decreases.  These difficulties
are {\small \bf most obvious among the intrinsically faint objects}
(presumably low in mass) and {\small \bf the objects depleted in
dust}.  In fact, the only two Oort Cloud comets between 2000 and
2017 observed to have survived in orbits with perihelion distance
not exceeding 0.3~AU, C/2006~P1 and C/2011~L4, were both very bright
(and presumably massive) and dust-rich objects.  It is proposed that
massive comets have a greater chance to survive because of their
substantial reservoirs of ice, whose sublimation is instrumental in
keeping the surface temperature under control; and that dust-rich
comets are able to better protect the surface of the nucleus by
saturating the atmosphere with microscopic dust to the point of
making it optically thick while close to the Sun.

By contrast, nearly all {\small \bf other long-period comets}
investigated in this paper {\small
\bf survive perihelion}, perhaps because of the presence
of a protective mantle of sintered dust on the surface of their
nuclei.  The process of sudden disappearance is among these (as well
as short-period) comets restricted to short-lived companions of the
split comets (see Section~10 of Sekanina \& Kracht 2018) and to
other isolated cases (e.g., 3D/Biela, 5D/Brorsen, 20D/Westphal), all
of which are independent of the heliocentric distance at perihelion.
These topics are outside the scope of this paper.  So is the problem
of disappearance of the dwarf sungrazers of the Kreutz system, which
sublimate away under extremely high temperatures.  

The perihelion survival limit for long-period comets, proposed by
Bortle (1991) as the minimum absolute brightness that the comet
has to have in order to survive, turned out to be too low:\ about
one third of the Oort Cloud comets investigated in this study ---
in both the 2000--2017 and 1989--1999~sets --- that were predicted
by this rule to survive did in fact perish. In an effort to remedy
this problem, I propose a {\small \bf synoptic index {\boldmath
$\Im_{\rm surv}$} as a new perihelion survival predictor},
but this issue should by no means be considered closed.  A larger
and more accurate data base should in the future allow a better
understanding of the nucleus' properties of Oort Cloud comets in
general and the forces that determine their surviving or perishing
near the Sun in particular. 

One comet, C/2016 U1, appears to have survived, defying both Bortle's
rule and the $\Im_{\rm surv}$ index.  I question whether this comet
was at all an Oort Cloud comet rather than an interloper that was
masquerading as a member thanks to the planetary perturbations that
in the previous return to the Sun modified its orbit by increasing
its aphelion toward the Oort Cloud; in fact, this comet's Oort Cloud
membership is already open to doubt on account of the fairly large
uncertainty of its computed original semimajor axis.

In closing, I remark that the proposed morphological similarity of `Oumuamua
and its parent comet with the Oort Cloud comets opens up intriguing
possibilities for research of interstellar objects.   The investigation
of perihelion survival of the Oort Cloud comets provides a remarkable
insight into the nature of their nuclei, especially in the absence of
their closeup imaging.  It is hoped that this paper will stimulate
increased interest in these and the other examined issues. \\
%
%

This research was carried out at the Jet Propulsion Laboratory, California
Institute of Technology, under contract with the National Aeronautics and
Space Administration.\\[0.3cm]
\begin{center}
{\bf APPENDIX A} \\[0.2cm]
{\rm ACCOMMODATION LIMIT, SYNOPTIC INDEX, AND EQUIVALENT PERIHELION
 DISTANCE\\FOR STRONGLY HYPERBOLIC ORBITS} \\
\end{center}
As the expression for the rate of increase in the{\vspace{-0.04cm}} solar
flux, $\dot{\cal F}_{\mbox{\tiny \boldmath $\odot$}}$, depends in part on the
object's orbital velocity, Equation~(6) applies strictly only to parabolic
motion, with high accuracy also to Oort Cloud comets.  It does not apply to
objects in strongly hyperbolic orbits, such as 1I/`Oumuamua or its parent
comet.  Because of the higher orbital velocity, the rate of increase in the
solar flux at a heliocentric distance $r$ along a hyperbolic{\vspace{-0.04cm}}
orbit, $(\dot{\cal F}_{\mbox{\tiny \boldmath $\odot$}})_{\rm h}(r;q,e)$, {\it
exceeds\/} the rate along a parabolic orbit{\vspace{-0.08cm}} of equal perihelion
distance, $(\dot{\cal F}_{\mbox{\tiny \boldmath $\odot$}})_{\rm p}(r;q)$.  The
same applies to the peak rates.  To account for this difference in the
expression for the accommodation limit ${\cal A}$ and in a plot of $H_0$
against $q$, as well as in the formula for the synoptic index $\Im_{\rm
surv}$, one should replace the perihelion distance $q$ of the hyperbolic
orbit with a perihelion distance $q^\prime$ of an equivalent parabolic
orbit (\mbox{$q^\prime < q$}), so that the peak rates of increase in the
solar flux match each other,
\begin{equation}
(\dot{\cal F}_{\mbox{\tiny \boldmath $\odot$}})_{\rm h,max}(q,e) =
 (\dot{\cal F}_{\mbox{\tiny \boldmath $\odot$}})_{\rm p,max}(q^\prime) .
\end{equation}
The general expression for the radial velocity is
\begin{equation}
\dot{r} = \frac{ke \sin u}{\sqrt{q (e \!+\! 1)}},
\end{equation}
where $k$ is the Gaussian gravitational constant and $u$ is the true
anomaly (\mbox{$u < 0$} before perihelion).  Providing $u$ in terms
of heliocentric distance, the expression for the rate of increase in
the solar flux along a hyperbolic orbit has a form
\begin{equation}
(\dot{\cal F}_{\!\mbox{\tiny \boldmath $\odot$}})_{\rm h}(r;q,e) =
 2k \, \frac{{\cal F}_0}{r^3} \, \sqrt{\frac{e \!-\! 1}{q}} \left[ 1 \!+\!
 \frac{q}{r} \, \frac{2}{e \!-\! 1} \!-\! \left( \:\!\! \frac{q}{r} \:\!\!
 \right)^{\! 2} \:\!\! \frac{e \!+\! 1}{e \!-\! 1} \right]^{\!\frac{1}{2}} \!\!\!.
\end{equation}
The rate of increase in the solar flux along a hyperbolic orbit
reaches a peak of
\begin{equation}
(\dot{\cal F}_{\mbox{\tiny \boldmath $\odot$}})_{\rm h,max}(q,e) =
 k{\cal F}_0 q^{-\frac{7}{2}} f^{-\frac{7}{2}} \sqrt{g}
\end{equation}
at
\begin{equation}
r_{\rm h,max}(q,e) = f q,
\end{equation}
where
\begin{equation}
f = \frac{\sqrt{1 + 48 e^2} - 7}{6 (e \!-\! 1)}
\end{equation}
and
\begin{equation}
g = 8 \! \left(\!1\!+\!\frac{e \!-\! 1}{2} f \!-\! \frac{e \!+\! 1}{2f} \!\right)\!.
\end{equation}
According to Equation\,(20),\,the perihelion distance~of~the equivalent parabolic
orbit, $q^\prime$, is determined by equating the expressions (4) at $q^\prime$
and (23) at $q$.  The result is
%
\begin{equation}
q^\prime = \frac{7f}{8} g^{-\frac{1}{7}} q.
\end{equation}
First, in the limiting~parabolic~case,~one~has~\mbox{$e \rightarrow 1$}~and~by
L'Hospital's Rule \mbox{$f \rightarrow \frac{8}{7}$},~{\vspace{-0.05cm}}which~\mbox{implies}~\mbox{$g \rightarrow 1$}~and,~from Equation~(27),
\mbox{$q^\prime \rightarrow q$}, as
expected.  Since~\mbox{`Oumuamua's} elements are \mbox{$q = 0.255$ AU},~\mbox{$e =
1.20$},~one~finds~for~its {\it equivalent\/} perihelion
distance \mbox{$q^\prime = 0.973 \:\!q = 0.248$ AU}, thereby correcting $q$ for the
effect of the hyperbolic~excess.  The
deviation from the nominal perihelion distance is trivial --- less
than 3~percent --- and can safely~be~neglected in the~first approximation.\\[0.1cm]
\begin{center}
{\bf APPENDIX B} \\[0.2cm]
{\rm AVERAGING PROJECTED AREA OF SUNLIT\\FRACTION OF OBLATE SPHEROID'S
 SURFACE\\OVER ALL DIRECTIONS} \\
\end{center}
In compliance with the conclusions of Section 9 I assume that `Oumuamua's figure
is approximated by an oblate spheroid, whose semiaxes are $a$, $a$, and $b$
(\mbox{$b \!<\! a$}) and whose maximum projected area in the
equatorial plane equals $\pi a^2$.  Let \mbox{$\gamma = a/b$} and the Sun make
an angle of $\theta$ with the normal to the equatorial plane, as shown in
Figure~B.1.  The other angle, $\phi$, is reckoned along the equatorial plane.
The task is to determine the projected area of the fraction of the spheroid's
surface that is sunlit (and therefore subjected to solar radiation pressure)
when averaged over all directions.  From Figure~B.1 it follows that at an angle
$\theta$ the projected sunlit area is an ellipse whose semimajor axis is $a$,
{\vspace{-0.1cm}}semiminor axis is $z(\theta)$, and the cross section
perpendicular to $\widehat{\rm SC}$ {\vspace{-0.1cm}} is $\pi a z(\theta)$.  Here
\mbox{$z(\theta) = \widehat{\rm CE}$} and point E lies on a tangent to the
{\vspace{-0.1cm}}spheroid that passes through point P$(x,y)$ and is parallel
to the direction $\widehat{\rm SC}$.  The tangent intersects the normal to
the equatorial plane at point F, delimiting an ordinate $\bar{y}$ and
making angle $\theta$ with the normal.

\begin{figure}[t]
\vspace{-6.1cm}
\hspace{0.78cm}
\centerline{
\scalebox{0.77}{
\includegraphics{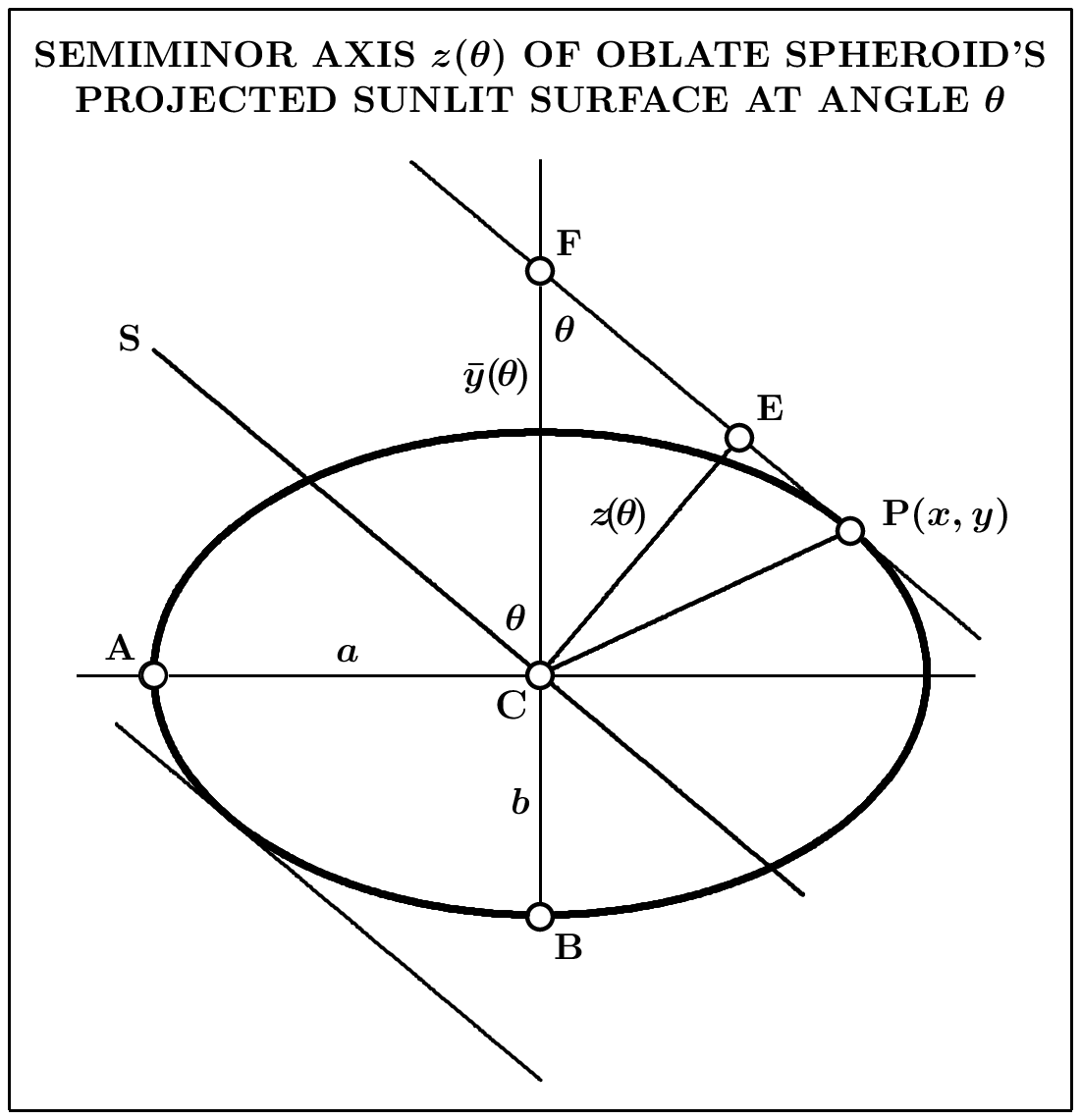}}}

\vspace{-7.6cm}
\parbox{8.6cm}{\footnotesize {\bf Figure B.1.} Determination of the semiminor
axis $z(\theta)$ of an oblate spheroid's projected sunlit surface at an
{\vspace{-0.07cm}}angle $\theta$.  The spheroid, centered at C and of
{\vspace{-0.07cm}}semimajor axis \mbox{$a = \widehat{\rm AC}$} and semiminor
axis \mbox{$b = \widehat{\rm BC}$}, is viewed from a point in its equatorial
{\vspace{-0.07cm}}plane in a direction perpendicular to incoming sunlight
{\vspace{-0.07cm}}along the line $\widehat{\rm SC}$.  The tangent
$\widehat{\rm FP}$ to the spheroid, parallel to $\widehat{\rm SC}$, delimits
{\vspace{-0.07cm}}the ordinate \mbox{$\bar{y} = \widehat{\rm CF}$} and the
semiminor axis \mbox{$z(\theta) = \widehat{\rm CE}$} of the sunlit area,
whose projected area is $\pi a z(\theta)$.}

{\vspace{0.7cm}}
\end{figure}

To determine the length of $z(\theta)$, I begin with an equation of the
spheroid in projection onto a plane perpendicular to the equatorial plane,
with the origin C$(x_0,y_0)$ at \mbox{$x_0 = 0, \, y_0 = 0$},
\begin{equation}
\frac{x^2}{a^2} + \frac{y^2}{b^2} = 1.
\end{equation}
Introducing \mbox{$\gamma = a/b$} (\mbox{$\gamma > 1$}), one has for the
tangent to the ellipse at point P$(x,y)$
\begin{table}[t]
\vspace{-4.2cm}
\hspace{4.2cm}
\centerline{
\scalebox{1}{
\includegraphics{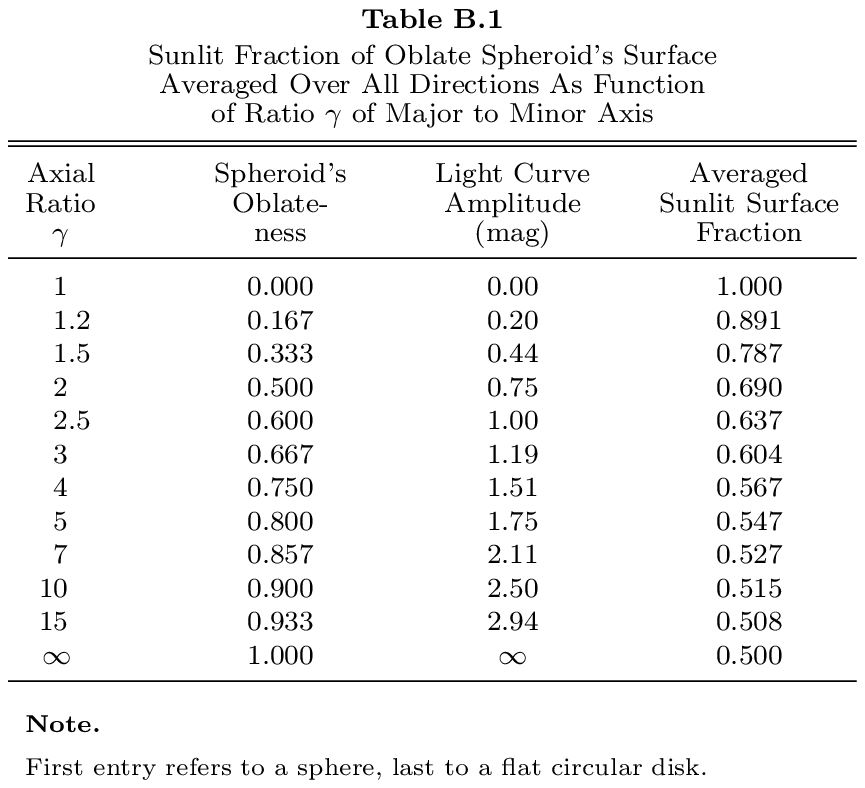}}}
\vspace{-17.5cm}
\end{table}
\begin{equation}
\frac{dy}{dx} = -\frac{x}{\gamma^2 y} = \tan (90^\circ \!+\! \theta) =
  -\cot \theta,
\end{equation}
so that
\begin{equation}
x = \gamma^2 y \cot \theta.
\end{equation}
To separate $x$ from $y$, I insert for $x$ from Equation~(30) into Equation~(28),
which gives
\begin{equation}
y = \frac{a \sin \theta}{\gamma \sqrt{ 1 + (\gamma^2 \!-\! 1) \cos^2 \theta}} \, ,
\end{equation}
and back from (30),

{\vspace{-0.2cm}}
\begin{equation}
x = \frac{ a \gamma \cos \theta}{\sqrt{1 + (\gamma^2 \!-\! 1) \cos^2 \theta}} \,.
\end{equation}
The ordinate $\bar{y}$ is computed from
\begin{equation}
\bar{y} = y + x \cot \theta ,
\end{equation}
allowing one to find for $z(\theta)$ an expression,
\begin{equation}
z(\theta) = \bar{y} \sin \theta = \frac{a}{\gamma} \sqrt{ 1 + (\gamma^2
  \!-\! 1) \cos^2 \theta)}.
\end{equation}
The projected area of the sunlit fraction of the spheroid's surface, averaged
over the hemisphere, $\langle X \rangle$, is now derived by integrating over
all angles $\theta$ from 0 to $\frac{1}{2} \pi$ and over all angles $\phi$
[which $z(\theta)$ is independent of] from 0 to 2$\pi$:
\begin{equation}
\int_{0}^{2\pi} \!\!\!\!d\phi \!\int_{0}^{\frac{\pi}{2}} \!\!\langle X \rangle
 \sin \theta \, d \theta = \!\!\int_{0}^{2\pi} \!\!\!\! d \phi \!
 \int_{0}^{\frac{\pi}{2}} \!\!\! \pi a z(\theta) \sin \theta \, d \theta
\end{equation}
with $z(\theta)$ from Equation~(34).  Employing a substitution \mbox{$\psi =
 \cos \theta$}, Equation~(35) becomes
to
\begin{eqnarray}
\langle X \rangle & = & \frac{\pi a^2}{\gamma} \!\! \int_{0}^{1} \!\!\!\sqrt{
 1 + (\gamma^2 \!-\! 1) \psi^2} \, d\psi \nonumber \\[0.15cm]
 & = & \frac{\pi a^2}{2} \!\left[ 1 + \frac{\ln(\gamma \!+\! \sqrt{\gamma^2
 \!-\! 1})} {\gamma \sqrt{\gamma^2 \!-\! 1}} \right] \!,
 \end{eqnarray}
which is the resulting relation.  Since \mbox{$\pi a^2$} is the maximum projected
area of the spheroid, this formula indicates that the averaged projected
{\vspace{-0.03cm}}area $\langle X \rangle$ is never smaller than $\frac{1}{2}$
the maximum.  It is equal to this value when \mbox{$\gamma \rightarrow \infty$},
i.e., for a flat circular disk.  At the other extreme,{\vspace{-0.04cm}}
$\langle X \rangle$ equals \mbox{$\pi a^2$} when \mbox{$\gamma = 1$}, i.e.,
for a sphere, as expected.  The averaged projected area of all spheroids
(\mbox{$1 < \gamma < \infty$}) is larger than one half of, and less than,
the maximum area.

The variation of the averaged sunlit surface fraction, $\langle X \rangle/\pi a^2$,
with the ratio $\gamma$ is listed in Table B.1; also included are the oblateness,
$(\gamma \!-\! 1)/\gamma$, of the spheroid and the amplitude of the light curve.
 For \mbox{$6 \leq \gamma \leq 11$}, the range of axial ratios relevant to the
light curve of `Oumuamua, the averaged sunlit surface fraction is restricted to
0.52\,$\pm$\,0.01. \\
\begin{center}
{\footnotesize REFERENCES}
\end{center}
\vspace{-0.25cm}
\begin{description}
{\footnotesize
\item[\hspace{-0.3cm}]
A'Hearn, M.\ F., Schleicher, D.\ G., Millis, R.\ L., et al.\ 1984, AJ, 89,{\linebreak}
  {\hspace*{-0.6cm}}579
\\[-0.57cm]
\item[\hspace{-0.3cm}]
A'Hearn, M.\ F., Millis, R.\ L., \& Schleicher, D.\ G.\ 1995, Icarus,
  118,{\linebreak}
  {\hspace*{-0.6cm}}223
\\[-0.57cm]
\item[\hspace{-0.3cm}]
Bonev, T., Jockers, K., Petrova, E., et al.\ 2002, Icarus, 160, 419
\\[-0.57cm]
\item[\hspace{-0.3cm}]
Borisov,\,G., Bonev,\,T., Iljev,\,I., \& Stateva,\,I.\ 2012,
  Bulg.~Astron.~J.,{\linebreak}
 {\hspace*{-0.6cm}}18, 47
\\[-0.57cm]
\item[\hspace{-0.3cm}]
Bortle, J.\ E.\ 1991, Int. Comet Quart., 13, 89
\\[-0.57cm]
\item[\hspace{-0.3cm}]
Camilleri, P.\ 1994, IAUC 5995
\\[-0.57cm]
\item[\hspace{-0.3cm}]
Curdt, W., Boehnhardt, H., \& Vincent, J.-B.\ 2014, A\&A, 567, L1{\hspace{0.2cm}}
\\[-0.57cm]
\item[\hspace{-0.3cm}]
Drahus, M., Guzik, P., Waniak, W., et al.\ 2018, Nature Astron.,~2,{\linebreak}
  {\hspace*{-0.6cm}}407
\\[-0.57cm]
\item[\hspace{-0.3cm}]
Farnham, T.\ L., Schleicher, D.\ G., Woodney, L.\ M., et al.\ 2001,{\linebreak}
 {\hspace*{-0.6cm}}Science, 292, 1348
\\[-0.57cm]
\item[\hspace{-0.3cm}]
Ferr\'{\i}n, I.\ 2014, MNRAS, 442, 1731
\\[-0.57cm]
\item[\hspace{-0.3cm}]
Fink, U., \& Rubin, M.\ 2012, Icarus, 221, 721
\\[-0.57cm]
\item[\hspace{-0.3cm}]
Gehrz, R.\ D.\ 1992, IAUC 5482
\\[-0.57cm]
\item[\hspace{-0.3cm}]
Gilmore, A.\ C.\ 1994, IAUC 6004
\\[-0.57cm]
\item[\hspace{-0.3cm}]
Green, D.\ W.\ E.\ 2004, IAUC 8346
\\[-0.57cm]
\item[\hspace{-0.3cm}]
Green, D.\ W.\ E.\ 2005, IAUC 8540
\\[-0.57cm]
\item[\hspace{-0.3cm}]
Guido, E., Sostero, G., \& Howes, N.\ 2011, CBET 2876
\\[-0.57cm]
\item[\hspace{-0.3cm}]
Hale, A.\ 1992, IAUC 5496
\\[-0.57cm]
\item[\hspace{-0.3cm}]
James, N.\ 2017, JBAA, 127, 132
\\[-0.57cm]
\item[\hspace{-0.3cm}]
Jehin, E., Boehnhardt, H., Sekanina, Z., et al.\ 2002, Earth Moon{\linebreak}
  {\hspace*{-0.6cm}}Plan., 90, 147
\\[-0.57cm]
\item[\hspace{-0.3cm}]
Jewitt, D., Luu, J., Rajagopal, J., et al.\ 2017, ApJ, 850, L36
\\[-0.57cm]
\item[\hspace{-0.3cm}]
Jorda, L., Hainaut, O., \& Smette, A.\ 1995, Plan.\ Space Sci.,~43,~737
\\[-0.57cm]
\item[\hspace{-0.3cm}]
Keane, J.\,V., Milam, S.\,N., Coulson, I.\,M., et al.\,2016, ApJ, 831,~207
\\[-0.57cm]
\item[\hspace{-0.3cm}]
Kidger, M.\ R.\ 2002, Earth Moon Plan., 90, 157
\\[-0.57cm]
\item[\hspace{-0.3cm}]
Kidger, M.\ R., Altieri, B., M\"{u}ller, T., \& Gracia, J.\ 2016, Earth{\linebreak}
  {\hspace*{-0.6cm}}Moon Plan., 117, 101
\\[-0.57cm]
\item[\hspace{-0.3cm}]
Kimura, H., \& Mann, I.\ 1998, J.\ Quant. Spec.\ Rad.\ Transf.,~60,~425
\\[-0.57cm]
\item[\hspace{-0.3cm}]
Knight, M.\ M., \& Battams, K.\ 2014, ApJ, 782, L37
\\[-0.57cm]
\item[\hspace{-0.3cm}]
Kobayashi, T.\ 1992, IAUC 5496
\\[-0.57cm]
\item[\hspace{-0.3cm}]
Korsun, P., Kulyk, I., \& Velichko, S.\ 2012, Plan.\ Space Sci., 60, 255
\\[-0.57cm]
\item[\hspace{-0.3cm}]
Kronk, G.\ W.\ 2003, Cometography:\ A Catalog of Comets,{\linebreak}
 {\hspace*{-0.6cm}}Volume 2:\ 1800--1899 (Cambridge, UK:\ Cambridge
 University{\linebreak}
 {\hspace*{-0.6cm}}Press), 900pp
\\[-0.57cm]
\item[\hspace{-0.3cm}]
Li, J., \& Jewitt, D.\ 2015, AJ, 149, 133
\\[-0.57cm]
\item[\hspace{-0.3cm}]
Machholz, D.\ E.\ 1994, J.\ Assoc.\ Lunar Plan.\ Obs., 37, 171
\\[-0.57cm]
\item[\hspace{-0.3cm}]
Machholz, D.\ E.\ 1996, J.\ Assoc.\ Lunar Plan.\ Obs., 39, 71
\\[-0.57cm]
\item[\hspace{-0.3cm}]
M\"{a}kinen, J.\ T.\ T., Bertaux, J.-L., Combi, M.\ R., \&
 Qu\'emerais,~E.{\linebreak}
 {\hspace*{-0.6cm}}2001, Science, 292, 1326
\\[-0.57cm]
\item[\hspace{-0.3cm}]
Marcus, J.\ N.\ 2007, Int.\ Comet Q., 29, 39
\\[-0.57cm]
\item[\hspace{-0.3cm}]
Marsden, B.\ G.\ 1990, AJ, 99, 1971
\\[-0.57cm]
%
%
%
\item[\hspace{-0.3cm}]
Marsden, B.\ G.\ 2000a, MPEC 2000-O07
\\[-0.57cm]
\item[\hspace{-0.3cm}]
Marsden, B.\ G.\ 2000b, MPC 28557
\\[-0.57cm]
\item[\hspace{-0.3cm}]
Marsden, B.\ G., \& Williams, G.\ V.\ 2008, Catalogue of Cometary{\linebreak}
 {\hspace*{-0.6cm}}Orbits 2008, 17th ed.\ (Cambridge, MA:\ IAU Minor
 Planet Cen-{\linebreak}
 {\hspace*{-0.6cm}}ter/Central Bureau for Astronomical Telegrams), 195pp
\\[-0.57cm]
\item[\hspace{-0.3cm}]
Marsden, B.\ G., Sekanina, Z., \& Yeomans, D.\ K.\ 1973, AJ, 78,~211
\\[-0.57cm]
\item[\hspace{-0.3cm}]
Marsden, B.\ G., Sekanina, Z., \& Everhart, E.\ 1978, AJ, 83, 64
\\[-0.57cm]
\item[\hspace{-0.3cm}]
Mattiazzo, M.\ 2003, IAUC 8250
\\[-0.57cm]
\item[\hspace{-0.3cm}]
Mattiazzo, M., \& McNaught, R.\ H.\ 2011, CBET 2801
\\[-0.57cm]
\item[\hspace{-0.3cm}]
Micheli, M., Farnocchia, D., Meech, K.\ J., et al.\ 2018, Nature, 559,{\linebreak}
 {\hspace*{-0.6cm}}223
\\[-0.57cm]
\item[\hspace{-0.3cm}]
Moreno, F., Pozuelos, F., Aceituno, F., et al.\ 2014, ApJ, 791, 118
\\[-0.57cm]
\item[\hspace{-0.3cm}]
Nakano, S.\ 1996, MPC 28557
\\[-0.57cm]
%
%
\item[\hspace{-0.3cm}]
Opitom, C., Jehin, E., Manfroid, J., \& Gillon, M.\ 2013, CBET 3433
\\[-0.57cm]
\item[\hspace{-0.3cm}]
Osip, D., Schleicher, D., \& Campins, H.\ 1993, Lunar Plan.\ Inst.,{\linebreak}
  {\hspace*{-0.6cm}}Contr.\ 810, 241
\\[-0.57cm]
\item[\hspace{-0.3cm}]
Protopapa, S., Kelley, M.\ S.\ P., Yang, B., et al.\ 2018, ApJ,~862,~L16
\\[-0.57cm]
%
%
\item[\hspace{-0.3cm}]
Rosenbush, V.\ K., Velichko, F.\ P., Kiselev, N.\ N., et al.~2006,~Solar{\linebreak}
 {\hspace*{-0.6cm}}Syst.\ Res., 40, 290
\\[-0.57cm]
%
%
%
\item[\hspace{-0.3cm}]
Schleicher, D.\ G., \& Osip, D.\ J.\ 1990, IAUC 4983
\\[-0.57cm]
\item[\hspace{-0.3cm}]
Scotti, J.\ V.\ 1994, IAUC 6004
\\[-0.57cm]
\item[\hspace{-0.3cm}]
Sekanina, Z.\ 1984, Icarus, 58, 81
\pagebreak
\item[\hspace{-0.3cm}]
Sekanina, Z.\ 2002, Int. Comet Quart., 24, 223
\\[-0.57cm]
\item[\hspace{-0.3cm}]
Sekanina, Z.\ 2005, IAUC 8545
\\[-0.57cm]
\item[\hspace{-0.3cm}]
Sekanina, Z.\ 2011, CBET 2876
\\[-0.57cm]
\item[\hspace{-0.3cm}]
Sekanina, Z.\ 2019, eprint arXiv:1901.08704
\\[-0.57cm]
%
%
\item[\hspace{-0.3cm}]
Sekanina, Z., \& Kracht, R.\ 2014, eprint arXiv:1404.5968
\\[-0.57cm]
\item[\hspace{-0.3cm}]
Sekanina, Z., \& Kracht, R.\ 2016, ApJ, 823, 2
\\[-0.57cm]
\item[\hspace{-0.3cm}]
Sekanina, Z., \& Kracht, R.\ 2018, eprint arXiv:1812.07054
\\[-0.57cm]
\item[\hspace{-0.3cm}]
Sekanina, Z., Jehin, E., Boehnhardt, H., et al.\ 2002, ApJ, 572, 679
\\[-0.57cm]
\item[\hspace{-0.3cm}]
Seki, T.\ 1992, IAUC 5496
\\[-0.57cm]
\item[\hspace{-0.3cm}]
Shanklin, J.\ D.\ 1997, JBAA, 107, 186
\\[-0.57cm]
\item[\hspace{-0.3cm}]
Shanklin, J.\ D.\ 1998, JBAA, 108, 305
\\[-0.57cm]
\item[\hspace{-0.3cm}]
Shanklin, J.\ D.\ 2001, JBAA, 111, 247
\\[1.5cm]
\item[\hspace{-0.3cm}]
Shanklin, J.\ D.\ 2009, JBAA, 119, 317
\\[-0.57cm]
%
%
\item[\hspace{-0.3cm}]
Tazaki, R., \& Nomura, H.\ 2015, ApJ, 799, 119 
\\[-0.57cm]
\item[\hspace{-0.3cm}]
Tomita, K.\ 1962, IAUC 1787
\\[-0.57cm]
\item[\hspace{-0.3cm}]
Tozzi,\,G.\,P., Boehnhardt,\,H., Hainaut,\,O.\,R., et al.\,2003,\,IAUC~8250
\\[-0.57cm]
\item[\hspace{-0.3cm}]
Trilling, D.\ E., Mommert, M., Hora, J.\ L., et al.\ 2018, AJ, 156,~261
\\[-0.57cm]
\item[\hspace{-0.3cm}]
Vsekhsvyatsky, S.\ K.\ 1958, Fizicheskie kharakteristiki komet.{\linebreak}
 {\hspace*{-0.6cm}}(Moscow:\ Gosud.\ izd-vo fiz.-mat.\ lit.); translated:\
 1964, Physical{\linebreak}
 {\hspace*{-0.6cm}}Characteristics of Comets, NASA TT-F-80 (Jerusalem:\
 Israel{\linebreak}
 {\hspace*{-0.6cm}}Program for Scientific Translations)
\\[-0.57cm]
\item[\hspace{-0.3cm}]
Weaver, H.\ A., Sekanina, Z., Toth, I., et al.\ 2001, Science, 292, 1249
\\[-0.57cm]
\item[\hspace{-0.3cm}]
Whipple, F.\ L.\ 1975, M\'em.\ Soc.\ Roy.\ Sci.\ Li\`{e}ge, S\'er.\ 6, 9, 101
\\[-0.64cm]
\item[\hspace{-0.3cm}]
Whipple, F.\ L.\ 1978, Moon \& Plan., 18, 343}
\vspace{1.38cm}
\end{description}
\end{document}